\documentclass[10pt, a4paper,twoside,american]{article}

\usepackage{geometry}
\usepackage{amsmath}
\usepackage{amssymb}

\usepackage{graphicx}

\usepackage{cite}

\usepackage{color} 
\usepackage{prettyref}

\usepackage[labelfont=bf,labelsep=period,justification=raggedright]{caption}

\makeatletter
\renewcommand{\@biblabel}[1]{\quad#1.}
\makeatother

\date{}

\usepackage[T1]{fontenc}
\usepackage[utf8]{inputenc}
\DeclareGraphicsExtensions{.pdf, .jpg, .tif}

\usepackage{fancyvrb}
\usepackage{fancyhdr}
\usepackage{url}

\usepackage{multirow}
\usepackage{colortbl}
\usepackage{tabularx}

\newcommand{\modelhdr}[3]{%
  \multicolumn{#1}{|l|}{%
    \color{white}\cellcolor[gray]{0.0}%
    \textbf{\makebox[0pt]{#2}\hspace{0.5\linewidth}\makebox[0pt][c]{#3}}%
  }%
}
\newcommand{\parameterhdr}[3]{%
  \multicolumn{#1}{|l|}{%
    \color{black}\cellcolor[gray]{0.8}%
    \textbf{\makebox[0pt]{#2}\hspace{0.5\linewidth}\makebox[0pt][c]{#3}}%
  }%
}
\newcommand{\nettypehdr}[3]{%
  \multicolumn{#1}{|l|}{%
    \color{black}\cellcolor[gray]{0.9}%
    \textit{\makebox[0pt]{#2}\hspace{0.5\linewidth}\makebox[0pt][c]{#3}}%
  }%
}

\begin{document}

% Title must be 150 characters or less
\begin{flushleft}
{\Large
\textbf{Frequency dependence of signal power and spatial reach of the local field potential}
}
% Insert Author names, affiliations and corresponding author email.
\\

\

Szymon Łęski$^{1, 2, \ast}$,
Henrik Lindén$^{1, 3}$,
Tom Tetzlaff$^{1, 4}$,
Klas H.~Pettersen$^{1}$,
Gaute T.~Einevoll$^{1}$

\

$^1$ Department of Mathematical Sciences and Technology, Norwegian University of Life Sciences, Ås, Norway
\\
$^2$ Department of Neurophysiology, Nencki Institute of Experimental Biology, Warsaw, Poland
\\
$^3$ Department of Computational Biology,
School of Computer Science and Communication,
Royal Institute of Technology (KTH),
Stockholm, Sweden
\\
$^4$ Institute of Neuroscience and Medicine (INM6), Research Center Jülich, Germany
\\
$\ast$ E-mail: s.leski@nencki.gov.pl
\end{flushleft}

% Please keep the abstract between 250 and 300 words
\section*{Abstract}

Despite its century-old use, the interpretation of local field potentials (LFPs), the low-frequency part of electrical signals recorded in the brain, is still debated. In cortex the LFP appears to mainly stem from transmembrane neuronal currents following synaptic activation, and obvious questions regarding the `locality' of the LFP are: What is the size of the signal-generating region, i.e., the spatial reach, around a recording contact? How far does the LFP signal extend outside a synaptically activated neuronal population? And how do the answers depend on the temporal frequency of the LFP signal? Experimental inquiries have given conflicting results, and we here pursue a modeling approach based on a well-established biophysical forward-modeling scheme incorporating detailed reconstructed neuronal morphologies in precise calculations of population LFPs including thousands of neurons.

The two key factors determining frequency dependence of LFP are (1) the spatial decay of the single-neuron LFP contribution and (2) the translation of synaptic input correlations into correlations between single-neuron LFP contributions. Both factors are seen to give low-pass filtering of the LFP signal power. For uncorrelated input only the first factor is relevant, and here a modest reduction in the spatial reach is observed for higher frequencies compared to the near-DC value ($\sim$~0~Hz) of about 200~$\mu\text{m}$.
Much larger frequency-dependent effects are seen when populations of pyramidal neurons receive correlated and spatially asymmetric inputs (basally or apically): the low-frequency ($\sim$~0~Hz) LFP power can here be an order of magnitude or more larger than the LFP power at, say, 60~Hz. Moreover, the low-frequency LFP components are found to have larger spatial reach and extend further outside the active population than high-frequency components. Our numerical findings are backed up by an intuitive simplified model for the generation of population LFP.

% Please keep the Author Summary between 150 and 200 words
% Use first person. PLoS ONE authors please skip this step.
% Author Summary not valid for PLoS ONE submissions.
\section*{Author Summary}

The first recording of electrical potential from brain activity was reported already in 1875, but still the interpretation of the signal is debated. To take full advantage of the new generation of microelectrodes with hundreds or even thousands of electrode contacts, an accurate quantitative link between what is measured and the underlying neural circuit activity is needed. Here we address the question of how the observed frequency dependence of recorded local field potentials (LFPs) should be interpreted. By use of a well-established biophysical modeling scheme, combined with detailed reconstructed neuronal morphologies, we find that correlations in the synaptic inputs onto a population of pyramidal cells may significantly boost the low-frequency components of the generated LFP. We further find that these low-frequency components may be less `local' than the high-frequency LFP components in the sense that (1) the size of signal-generation region of the LFP recorded at an electrode is larger and (2) that the LFP generated by a synaptically activated population spreads further outside the population edge due to volume conduction.

\section{Introduction}

The measurement of electrical potentials in the brain has a more than hundred year old history \cite{Caton:1875wl}. While the high-frequency part has been successfully used as a measure of spiking activity in a handful of surrounding neurons, the interpretation of the low-frequency part, the local field potential (LFP), has proved more difficult. Current-source density (CSD) analysis of multisite LFP recordings across well-organized layered neural structures such as cortex and hippocampus, was introduced in the 1950's \cite{Pitts:1952}. However, even if the CSD is a more local measure of neural activity than the LFP \cite{Nicholson:1975,Pettersen:2006gn,Leski:2007hw,Leski:2010bs,Leski:2011kr,Potworowski:2012jn}, the interpretation in terms of underlying activity in neural populations is inherently ambiguous \cite{Einevoll:2007fm,Pettersen:2012wy}.
Thus in many in vivo applications, for the example when investigating receptive fields in sensory systems, the LFP signal was discarded altogether.
The LFP signal has seen a revival in the last decade, however. This is due to the rapid development of new silicon-based microelectrodes now allowing for simultaneous recordings of LFP at tens or hundreds of contacts \cite{Normann:1999wi, Buzsaki:2004bi, Kipke:2008gg, Du:2011ik} (and availability of affordable high-capacity hard discs to store the data),
the realization among neuroscientists that the LFP offers a unique window into neural activity at the population level
\cite{Di:1990wq,Schroeder:1998uj,Henrie:2005dg,Einevoll:2007fm,Belitski:2008bq,
Mazzoni:2008cv,
Montemurro:2008uh,Kayser:2009ju,Mazzoni:2011ja,Szymanski:2011ib},
and the possibility of using the LFP signal in brain-machine interfaces
\cite{Mehring:2003dt,Andersen:2004en,Rickert:2005id,Markowitz:2011kt}.

To take full advantage of the opportunities offered by this new recording technology, a precise understanding of the link between the recorded LFP and the underlying neural activity is required. For example, two obvious questions regarding the `locality' of the LFP that need quantitative answers are: (1) What is the size of the signal-generating region, i.e., spatial reach, around a recording contact? (2) How far does the LFP signal extend outside an active population due to volume conduction? The first question has been addressed in several experimental studies, with resulting estimates for the spatial reach in cortex varying from a few hundred micrometers to several millimeters
\cite{Liu:2006cc,Berens:2008dq,Katzner:2009gp,Kreiman:2006dn,Xing:2009id,Kajikawa:2011bl}.
This large range in reported experimental estimates presumably reflects that the spatial reach depends strongly on the spatiotemporal properties of the underlying spiking network activity, in particular the level of correlations \cite{Linden:2011ck}. These critical network features will not only vary between the different brain regions and species studied, but also depend on the brain state.

In cortex, thousands of neurons contribute to the LFP, making the signal inherently difficult to interpret. Fortunately, the “measurement physics”, i.e., the biophysical link between neural activity and what is measured, is well understood: According to well-established volume-conductor theory \cite{RALL:1962ts,Pettersen:2012wy}, the recorded LFPs stem from appropriately weighted contributions from transmembrane currents in the vicinity of the electrode contact.  Building on pioneering work by Rall in the 1960's \cite{RALL:1962ts,Rall:1968}, a forward-modeling scheme incorporating detailed reconstructed neuronal morphologies in precise calculations of extracellular potentials, has been established \cite{Holt:1999wi} and used to explore both spikes
\cite{Holt:1999wi, Gold:2006ho, Gold:2007dy, Pettersen:2008p1158, Pettersen:2008iu}
and LFPs
\cite{Einevoll:2007fm, Pettersen:2008iu, Linden:2010p1109, Linden:2011ck, Gratiy:2011te}
generated by single neurons
\cite{Holt:1999wi, Gold:2006ho, Gold:2007dy, Pettersen:2008p1158, Linden:2010p1109}
and neural populations
\cite{Einevoll:2007fm, Pettersen:2008iu, Linden:2011ck}.
Unlike in experiments, this modeling scheme allows for a clear separation between volume conduction effects and effects of spatiotemporal variations in spiking network activity in determining population LFPs. In \cite{Linden:2011ck} it was used in a thorough investigation of the locality of LFP. It was found that the size of the LFP-generating region depends on the neuron morphology, the synapse distribution and correlations in synaptic activity. For uncorrelated activity, the LFP represents neurons in a small region (that is, a few hundred micrometers around the electrode contact), while in the case of correlated input the size of the generating region is determined by the spatial range of correlated synaptic activity and could thus be much larger.
Specifically, it was found that correlated synaptic inputs onto either the apical or basal dendrites of a population of pyramidal neurons could give orders of magnitude larger LFPs, and a much larger spatial reach, compared to the situations with (1) the same correlated input spread homogeneously over the neuronal dendrite or (2) similar uncorrelated synaptic inputs placed evenly or unevenly over the neurons.

As shown in \cite{Linden:2011ck}, the relative contributions to the population LFP from neurons at different distances from the electrode will depend on three factors: First, a single neuron close to the electrode will contribute more to the LFP than if it was placed further away.
Second, for a disc-like population, characteristic for a laminar population in a cortical column, it follows that with constant neuron density, the number of neurons located on a ring at a particular radial distance $r$ from the electrode will increase linearly with $r$. Third, with correlated synaptic inputs onto a neural population, the LFP contributions from different cells will also become correlated, or synchronized, and effectively boost the contributions to the LFP. The contributions from different rings of neurons will thus be determined by the interplay of these three factors. In \cite{Linden:2011ck} a simplified model for LFP generation based on these elements, (1) the decay of the single-neuron contribution with the distance from the electrode, (2) the population geometry, and (3) the correlation of LFP contributions from individual neural sources, was constructed. We found this simple model to not only give qualitative insight into the generation of population LFPs, but also quantitatively accurate predictions.

Strong frequency dependencies have been observed both in the tuning properties \cite{Liu:2006cc,Berens:2008dq} and information content \cite{Belitski:2008bq,Mazzoni:2011ja} of cortical LFPs. For example, the low-frequency LFP (less than 12~Hz) has been shown to carry complementary information to the gamma-range LFP (50-100~Hz) in V1 of macaque monkeys during naturalistic visual stimulation \cite{Mazzoni:2011ja}. To properly interpret such experiments, it is thus important to know how spatial reach of the LFP varies across frequencies and whether the biophysics of LFP signal generation boost some frequencies compared to others. The high-frequency LFP components are, for example, expected
to be more local than the low-frequency components due to intrinsic dendritic filtering \cite{Linden:2010p1109}, i.e.~due to the reduction of the (effective) current-dipoles with increasing frequency resulting from the capacitive properties of the dendritic membrane \cite{Pettersen:2012wy}.

In \cite{Linden:2011ck} we used the biophysical forward-modeling scheme to investigate the total population LFP, i.e., the total signal generated
across all frequencies. Here we use the same scheme to investigate both the distribution of the power of synaptically generated LFP between different frequency bands and the frequency dependence of the locality of the LFP signal. In terms of the latter, we study
the size of the signal-generating region (spatial reach) as well as the spatial extension of the LFP signal outside an active population --- for each frequency component separately.

\begin{figure}[!ht]
    \centering
        \includegraphics[width=\linewidth]{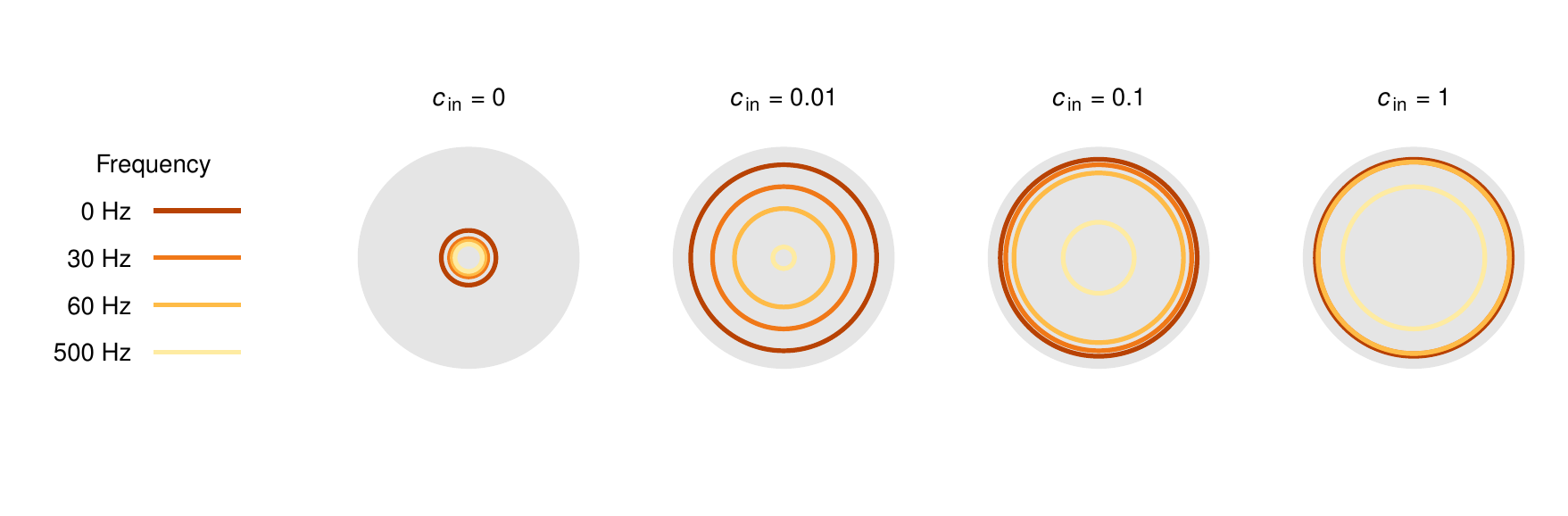}
    \caption{{\bf Spatial reach of different frequency components of LFP for different levels of synaptic input correlations $\mathbf{c}_\text{in}$.} Color lines denote parts of the whole population (gray, radius = 1~mm) which contribute 95\% of LFP amplitude at given frequency in the middle of the population, at the soma level. Results for layer-5 pyramidal cell with basal input.}
    \label{fig:sketch}
\end{figure}

We also use a frequency-resolved version of the simplified model developed in \cite{Linden:2011ck} to guide our investigation of this frequency dependence.
The population geometry (factor 2) does obviously not change with frequency. In contrast, the single-neuron LFP contribution (factor 1) decays faster for higher LFP frequencies \cite{Pettersen:2008p1158,Linden:2010p1109}, but an equally important factor turns out to be the frequency dependence  of the `correlation transfer', i.e., how correlations in the synaptic input are transferred to correlations between the single-neuron LFP contributions (factor 3). As an example, Figure 1 illustrates how the frequency-resolved spatial reach varies with the input correlation for a pyramidal population receiving basal synaptic inputs. We show that when the frequency dependencies of factors 1 and 3 are incorporated, the simplified model can still account well for the results obtained by comprehensive numerical investigations. To allow for direct use of the
simplified model in future applications, we here thus present and tabulate numerical results for the frequency dependence of these key factors for a variety of situations

The paper is organized as follows: first we describe our simulation setup, present
the simplified model of the population LFP, and review its ingredients.
Then we present detailed results of the simulations: we analyze the frequency content of the population LFP, the reach of different frequency components, and the decay of the signal outside of the population. Next we discuss the implications of our results for interpretation of electrophysiological data in terms of the underlying understanding the neural activity. Finally, in Methods we give details of the simulation setup and the mathematical model.

% Results and Discussion can be combined.
\section{Results}
\subsection{Simulations} % (fold)
\label{sub:simulation_setup}
\begin{figure}[!ht]
   \centering
       \includegraphics[width=11.9cm]{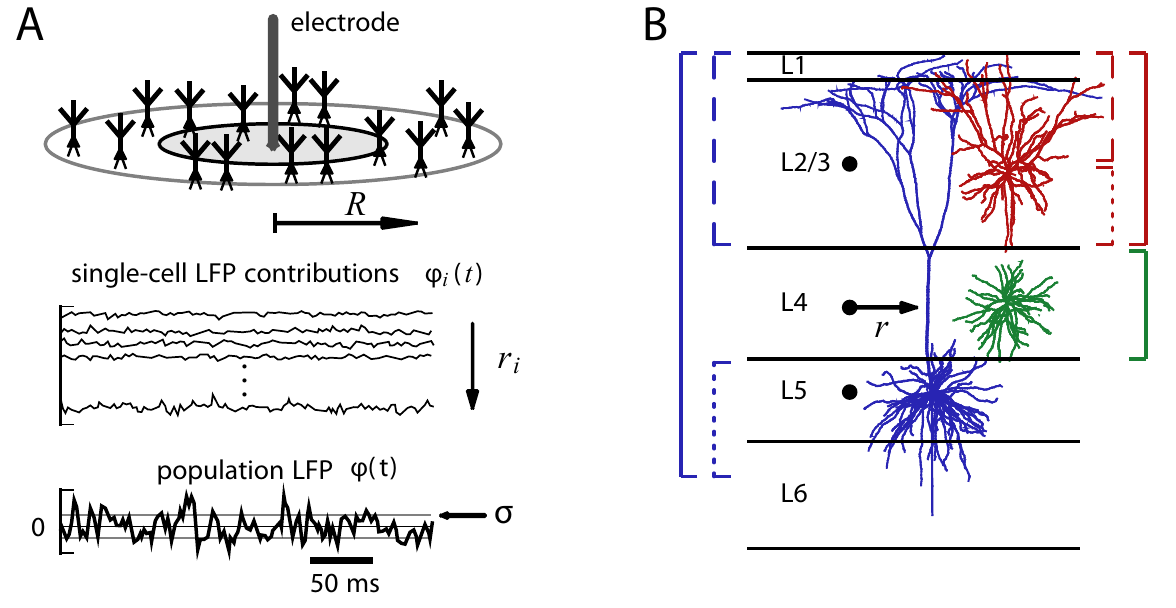}
   \caption{\textbf{Simulation setup.} A. Model cells are placed with constant area density on a disc of radius $R$, with
   the recording electrode at the population center, at soma level.
   The population LFP $\phi(t)$ is a sum of contributions $\phi_i(t)$ from cells at distances
   $r_i$. The dependence of its amplitude $\sigma(R)$  on the population radius R serves to define
   the spatial reach (see text).
   B. Reconstructed cell morphologies used in simulations, L3 pyramidal cell (red), L4 stellate cell (green), L5 pyramidal cell (blue), superimposed on the layers' boundaries.
   Electrode positions shown as black dots. Brackets mark the regions where synapses were distributed for either homogenous (solid brackets), apical (dashed brackets), or basal (dotted brackets) stimulation.
    Adapted from \cite{Linden:2011ck}.}
   \label{fig:simsetup}
\end{figure}
The fundamental formula relating neural activity to the generation of the extracellular electrical potential, including the LFP signal, is given by \cite{Holt:1999wi,Pettersen:2012wy}
\begin{equation}
    \label{eq:pointsource}
    \phi(\mathbf{r}, t) = \frac{1}{4\pi\sigma_{\text{cond}}}
    \sum_{i=1}^{n}        \frac{I_i(t)}{|\mathbf{r}-\mathbf{r}_i|},
\end{equation}
where $I_i$ denotes the transmembrane current in a neural compartment $i$ positioned at $\mathbf{r}_i$, and the extracellular conductivity is denoted by $\sigma_\text{cond}$. The transmembrane currents are calculated by means of standard multicompartmental modeling techniques with the simulation tool NEURON~\cite{Carnevale:2006vua}.

An essential part of the present work is the numerical simulation of the LFP in the center of a disc-like population
of cortical cells. The simulation setup is illustrated in Figure~\ref{fig:simsetup}. We consider a population of $N$=10000 cells distributed homogeneously on a planar disc with a radius of 1000~$\mu$m, Figure~\ref{fig:simsetup}A. The somas of the cells are positioned at the same depth, and the LFP is calculated at the soma level. In this setup we investigate how the LFP signal increases as contributions from more and more distant neurons are included, i.e., we study how the root mean square amplitude $\sigma$ of the population LFP $\phi(t)$ (obtained as a sum of single-cell contributions $\phi_i(t)$) depends on the radius $R$ of the subpopulation of cells included in the sum.

In the simulations we use three different morphologically-detailed cell models shown in Figure~\ref{fig:simsetup}B: the
layer-3 and layer-5 pyramidal cells, and the layer-4 stellate cells.
All neuron models are passive, i.e., without active conductances, and the extracellular signatures of action potentials (spikes) are thus not included.
For each class of pyramidal cells we consider three different spatial patterns of synaptic input: the synapses are placed either in the apical region only, in the basal region only, or evenly over the whole cell (Figure~\ref{fig:simsetup}B).
For the layer-4 stellate cells we consider only spatially homogeneous synaptic input, as these cells lack clearly defined dendritic regions.

The synaptic currents are modeled as $\alpha$-functions with a very short time constant ($\tau$=0.1~ms) to assure that no frequency filtering is imposed by the synapses themselves. In the frequency range considered in the present simulations (up to 500~Hz) each synaptic input current thus effectively corresponds to a $\delta$-function with a white (flat) power spectrum. With Poissonian spike statistics, which also implies a white power spectrum, the only frequency filtering in our simulation setup will come from the intrinsic dendritic filtering effect \cite{Pettersen:2008p1158,Linden:2010p1109} due to electrical properties of the cable and the summation of the single-neuron LFP contributions to form the population LFP.
For further details on the simulations we refer to the Methods section.

Note that we here for simplicity will refer to all calculated extracellular potentials as ``LFPs'' even if we consider frequencies as high as 500~Hz, i.e., frequencies often regarded to be outside the LFP band.

\subsection{Simplified model of population LFP} % (fold)
\label{sub:model}
To understand how the population signal emerges from single-cell contributions we use a simplified mathematical model, which is a frequency-resolved version of the model introduced in \cite{Linden:2011ck}.

We assume that the \emph{power spectral density (PSD)} of the contribution to the LFP from the $i$-th cell at given frequency can be factorized as
\begin{equation}
    \label{eq:factass}
    |\Phi_i(f)|^2 \approx
 \sigma_\xi^2(f) F_i^2(f),
\end{equation}
where
$\Phi_i$ is the PSD of the single-cell LFP,
$\sigma_\xi^2$ is the PSD of the input current, and $F_i(f)$ is the frequency-dependent \emph{shape function} of the $i$-th cell, which carries the information about how the root mean square amplitude of the signal at given frequency decays with distance.
Moreover, we assume that the shape function of each cell in the population can be replaced with a single, distance- and frequency-dependent function:
\[ F_i(f) = F(f, r_i), \]
that is, we assume that the shape function $F_i$ only depends on the frequency and the lateral distance $r_i$
from the recording electrode (Figure~\ref{fig:simsetup}B), and neglect variation in the single-neuron LFP contributions due
to other factors. For each particular morphology (layer-3/layer-4/layer-5) and synaptic stimulation pattern (homogeneous/apical/basal), the LFP contribution from each cell in the population is thus described with the function $F(f, r)$.
Note that for the special case of white-noise input (i.e., $\sigma_\xi^2(f) = \text{const.}$), the squared shape function $F(f,r_i)^2$ will be proportional to the PSD of the single-cell contribution to the LFP.

The summation of single-cell LFPs to the population signal depends on the correlation between the single-cell LFP contributions. In the case of \emph{uncorrelated} input this amounts to simply adding the variances of the single-cell LFPs. For a disc-like population of radius $R$ we thus obtain the following expression for the PSD of the signal at the center:
\begin{equation}
    \label{eqG0}
    G_0(f, R) = \sigma_\xi^2\sum_{r_i<R} \left| F(f, r_i)\right|^2 \to 2\sigma_\xi^2 \pi \rho \int_0^R r\left|F(f, r)\right|^2\text{d}r,
\end{equation}
where $\rho$ is the planar cell density.
On the other hand, if the single-cell LFPs are fully \emph{correlated}, the PSD of the signal is found by adding the single amplitudes, not variances, and we thus obtain
\begin{equation}
    \label{eqG1}
    G_1(f, R) = \sigma_\xi^2 \left|\sum_{r_i<R} F(f, r_i)\right|^2 \to \sigma_\xi^2(2\pi \rho)^2 \left|\int_0^R rF(f, r)\text{d}r\right|^2 .
\end{equation}

In our simulation setup the single-cell LFP contributions from two equidistant neurons (i.e., same $r_i$) are not identical even for $c_\text{in} = 1$: while the same spike trains are used to synaptically stimulate the cell, they will not in general activate an identical set of synapses (see Methods). Moreover, as we now work in the frequency domain, the correlation between single-cell contributions to the  LFP ($\phi_i$, $\phi_j$) is
naturally replaced by their \emph{coherence} ($\Phi_i^\ast \Phi_j / |\Phi_i||\Phi_j|$), which, in general, depends on the frequency.

If we approximate the LFP coherence between each pair of cells by the population-averaged LFP coherence $c_\Phi$, then the PSD is given by
\begin{equation}
    \label{eq:P}
   P(f, R) = [1-c_\Phi(f)] G_0(f, R) + c_\Phi(f) G_1(f, R)
\end{equation}
(see Methods for the full derivation of this formula). Note that the root mean square amplitude $\sigma$ of the signal (see Figure~\ref{fig:simsetup}) is related to the PSD $P(f, R)$ through
\begin{equation*}
    \sigma^2(R) = \int P(f,R) \text{d}f,
\end{equation*}
where the integration is between $f=0$~Hz and half the sampling frequency.

% subsection ingredients_of_the_model (end)
\subsection{Numerical evaluation of ingredients of simplified model} % (fold)

Equation \ref{eq:P} implies that any frequency dependence of the population LFP (for example, frequency dependence of the spatial reach) in general will result from the interplay of two separate effects: (1) frequency dependence of the single-cell shape functions $F(f,r)$ and (2) frequency dependence of the coherence $c_\Phi(f)$ between single-cell contributions to the population signal $P(f,R)$. These two effects are addressed next.

\begin{figure}[!ht]
    \centering
        \includegraphics[width=\linewidth]{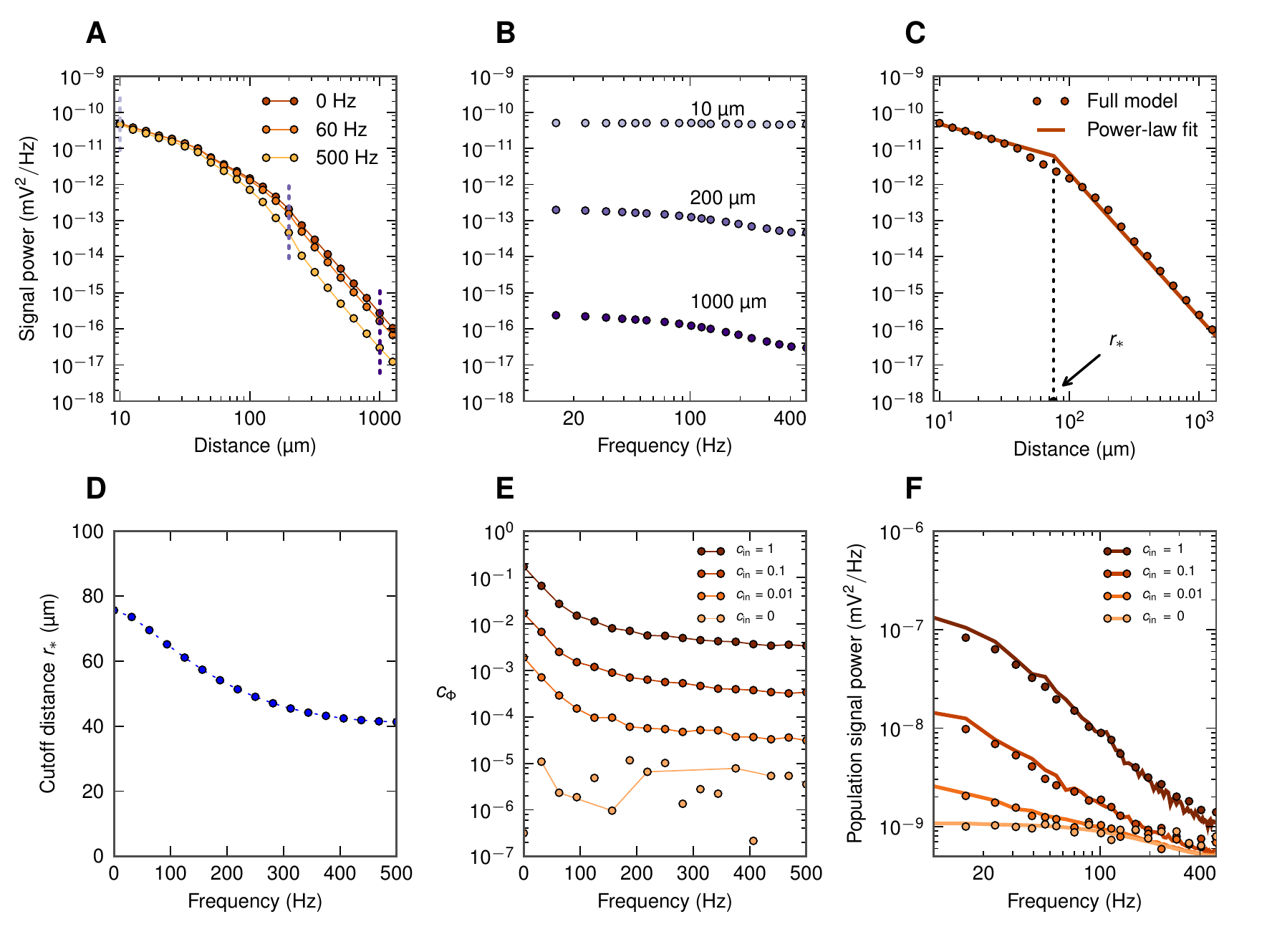}
    \caption{
    \textbf{Ingredients of the simplified LFP model for
     the layer-5 cell with basal synaptic input.}
      A. Spatial decay in lateral direction for the squared single-cell shape functions $|F(f,r)|^2$ for three different frequencies $f=0$, $60$ and $500$~Hz.
      B. Single-cell LFP spectra $|F(f,r)|^2$ for three different lateral distances from the soma
      (dotted vertical lines in A).
      C. Log-log plot of the squared near-DC ($\sim$~0~Hz) shape function $|F(0,r)^2|$ (dots) approximated
        by a piecewise-linear function with cutoff distance $r_\ast$ (line; see Eq.~\ref{Frast}).
      D. Frequency dependence of the cutoff distance $r_\ast(f)$.
      E. Population-averaged LFP coherence $c_{\Phi}$ for different input correlation levels $c_\text{in}$. Dots not connected with lines indicate that $|c_{\Phi}|$ is plotted in place of spurious negative values (see Methods).
      F. Power spectra $P(f,R)$ of the compound LFP ($R=1$~mm); dots correspond to simulation;  lines correspond to predictions from simplified model,
      Eq.~\ref{eq:P}, based on $r_\ast$ and $c_{\Phi}$ given in D and E, respectively.
    }
    \label{fig:fig_results_intro}
\end{figure}

\subsubsection{Frequency dependence of shape function}

The power of the extracellular potential from a single neuron decays when we move away from the cell, and the rate of the decay depends on the frequency of the signal.
In Figure~\ref{fig:fig_results_intro}A we have plotted squared shape functions $F(f,r)^2$ at the soma level for three selected frequency bands for the case with the layer-5 cell receiving basal synaptic stimulation.
We observe that the high-frequency LFP component decays faster with distance than the low-frequency component. This leads to the low-pass filtered power spectra seen in Figure~\ref{fig:fig_results_intro}B and is consistent with our previous observations of low-pass filtering in dendritic cables, i.e., the intrinsic dendritic filtering effect \cite{Pettersen:2008p1158,Linden:2010p1109}. To quantify this phenomenon we approximate the actual shape functions with simplified power-law shape functions. Specifically, at the soma level the amplitude of the single-cell LFP is, following \cite{Einevoll:2013wq}, modeled as:
\begin{equation}
    \label{Frast}
    F(f, r) = \begin{cases}
        F_0, &\text{ if }r<r_\epsilon,\\
        F_0 \sqrt{r_\epsilon / r}, &\text{ if }r_\epsilon \le r < r_\ast(f),\\
        F_0 \sqrt{r_\epsilon / r_\ast(f)}(r_\ast(f)/r)^2 &\text{ if }r \ge r_\ast(f),
    \end{cases}
\end{equation}
i.e., the shape function is approximated by $\propto r^{-1/2}$ close to the cell ($r<r_\ast$) and by $\propto r^{-2}$ (dipole) in the far-field regime ($r > r_\ast$). The parameter $r_\ast$ thus represents the \emph{cutoff distance} where the LFP contribution switches from the near-field ($F \propto r^{-1/2}$) to the far-field regime ($F \propto r^{-2}$), see fitted curve in Figure~\ref{fig:fig_results_intro}C.
This parametric representation of the shape function allows us to express the functions $G_0(f, R)$ and $G_1(f,R)$ (Equations~\ref{eqG0} and \ref{eqG1}) explicitly in terms of the cutoff distance $r_\ast$, see Methods for details. The observed reduction of $r_\ast$
with increasing frequency (Figure~\ref{fig:fig_results_intro}D) is intimately related to the corresponding reduction of the frequency-dependent electrotonic length constant in dendrites
\cite{Pettersen:2008p1158, Pettersen:2012wy}. In the example shown in Figure~\ref{fig:fig_results_intro}A the transition to dipole decay occurs closer to the cell for the high-frequency signal
(at about $\simeq 40\,\mu$m) than for the low-frequency components ($\simeq 80\,\mu$m).
\begin{figure}[!ht]
    \centering
        \includegraphics{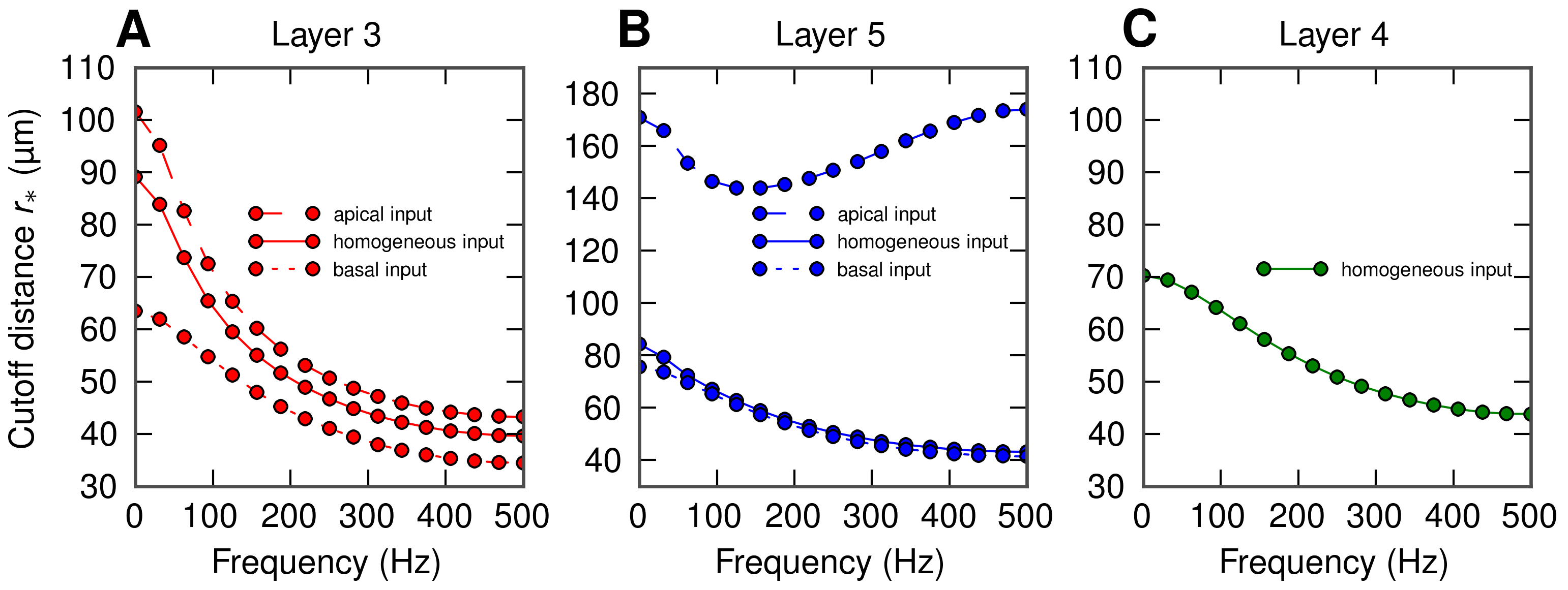}
    \caption{\textbf{Frequency dependence of the cutoff distance $r_\ast$ for all situations considered:}
 homogeneous (solid), apical (dashed) and basal synaptic input (dotted) applied to the layer-3 pyramidal cell (A), the layer-5 pyramidal cell (B), and the layer-4 stellate cell (C). Cell morphologies depicted in Figure~\ref{fig:simsetup}B.}
    \label{fig:rstars_popsim_all_long}
\end{figure}

In Figure~\ref{fig:rstars_popsim_all_long} we show the calculated cutoff distance $r_\ast$ for the various situations considered in the present paper involving the layer-3 pyramidal neuron (\ref{fig:rstars_popsim_all_long}A),
the layer-5 pyramidal neuron (\ref{fig:rstars_popsim_all_long}B), and the layer-4 stellate neuron (\ref{fig:rstars_popsim_all_long}C). For the pyramidal neurons we consider three spatial patterns
of synaptic inputs, that is homogeneous, only apical or only basal (Figure~\ref{fig:simsetup}).
All these combinations of cell morphology and stimulation pattern exhibit similar behavior as in our example (Figure~\ref{fig:fig_results_intro}): $r_\ast(f)$ decays with increasing frequency. The only exception is the layer-5 cell with apical input, where $r_\ast$ is very large, and also exhibits a minimum around $150$~Hz. This reflects that the geometry of this situation is unique, with the synaptic input positioned far above the soma level where the LFP is recorded. As a consequence the shrinkage of the current dipole with increasing frequency will be accompanied by a shift of the mean position of the current dipole in the apical direction \cite{Linden:2010p1109, Linden:2011ck}. The squared shape functions and the single-cell power spectra for the remaining situations (all apart from layer-5 cell with basal synaptic input) are shown in Figures~S2A,~B--S7A,~B.

\subsubsection{Frequency dependence of coherence}

The single-cell shape functions $F(f,r)$ alone are generally not sufficient to predict the population LFP. The missing component is $c_\Phi(f)$, the frequency-dependent population-averaged coherence between single-cell LFP contributions. This quantity can be estimated from population simulations, as described in detail in Methods, Equation~\ref{eq:cPhi}.
Coherence curves for different input correlation levels for our example (layer-5 cells receiving basal stimulation) are shown in Figure~\ref{fig:fig_results_intro}E.
The coherence $c_\Phi(f)$ is seen to be higher for low-frequency components. This may be understood on biophysical grounds by considering the dendritic morphology of the cell: for high-frequency synaptic input the return currents will be closer to the synaptic currents \cite{Pettersen:2008p1158}, and for the example in Figure~\ref{fig:fig_results_intro}E with basal stimulation of layer-5 pyramidal neurons, the resulting current dipoles will expectedly tend to be oriented randomly in space. However, for low-frequency input some of the synaptic input current will return through the apical dendrite \cite{Linden:2010p1109},
and the orientation of the effective current dipoles will be more similar between cells, leading to a higher coherence.

By combining the shape functions $F(f,r)$ with the LFP coherence $c_\Phi(f)$ in the simplified model (Equation~\ref{eq:P}) we can now obtain predictions for the population LFP. The resulting PSD for our example situation is shown in Figure~\ref{fig:fig_results_intro}F and is seen to be in excellent agreement with the simulation results. (See Figures~S2C--S7C for the results for the remaining combinations of cell type and synaptic input
patterns.)

\begin{figure}[!ht]
    \centering
        \includegraphics[width=\linewidth]{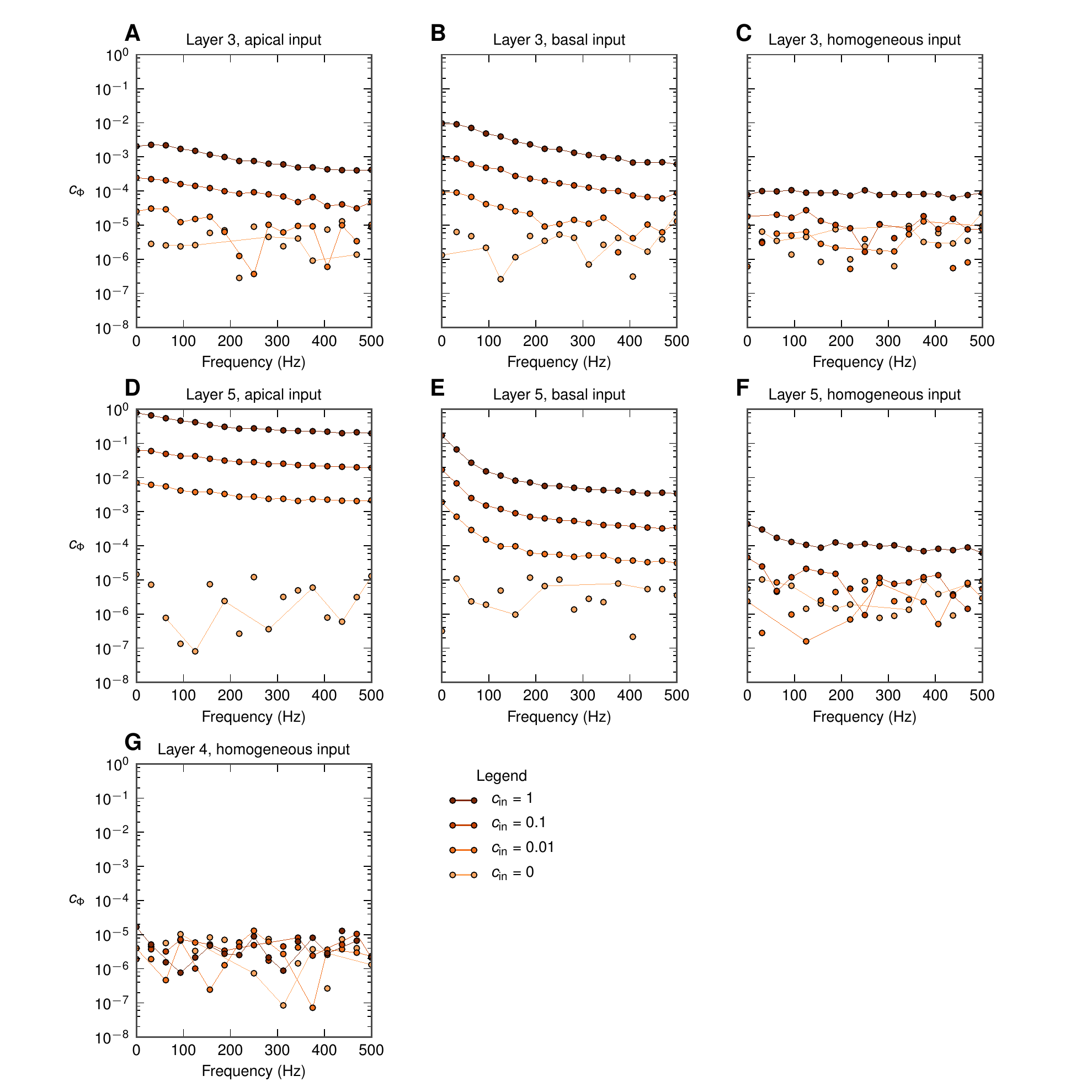}
    \caption{\textbf{Frequency dependence of the population-averaged LFP coherence $\mathbf{c}_{\Phi}$ for all situations considered. }  Dots not connected with lines indicate that $|c_{\Phi}|$ is plotted, see Methods. }
    \label{fig:phibars_popsim_all_long}
\end{figure}

In Figure~\ref{fig:phibars_popsim_all_long} we show the frequency dependence of the coherence $c_\Phi(f)$ for the same full set of seven situations as depicted in Figure~\ref{fig:rstars_popsim_all_long}. A first observation is that for pyramidal neurons (layer-3, layer-5) with asymmetric synaptic input (either only apical or only basal), decay of $c_\Phi(f)$ with increasing frequency is observed for all non-zero levels of input correlations $c_\text{in}$. This low-pass filtering effect is seen to be strongest for the layer-5 cell with basal input (Figure~\ref{fig:phibars_popsim_all_long}A, \ref{fig:phibars_popsim_all_long}B, \ref{fig:phibars_popsim_all_long}D, \ref{fig:phibars_popsim_all_long}E). However, when the same pyramidal neurons receive homogeneous synaptic inputs, the filtering effect is almost absent (Figure~\ref{fig:phibars_popsim_all_long}C, \ref{fig:phibars_popsim_all_long}F). In that respect it
resembles the situation with the stellate layer-4 cells receiving homogeneous synaptic input (Figure~\ref{fig:phibars_popsim_all_long}G) where $c_\Phi$  is essentially zero, implying that
the correlations in the synaptic input do not translate into correlations of the single-neuron LFP contributions.

% subsection overview (end)

%
\begin{figure}[!ht]
    \centering
        \includegraphics[width=\linewidth]{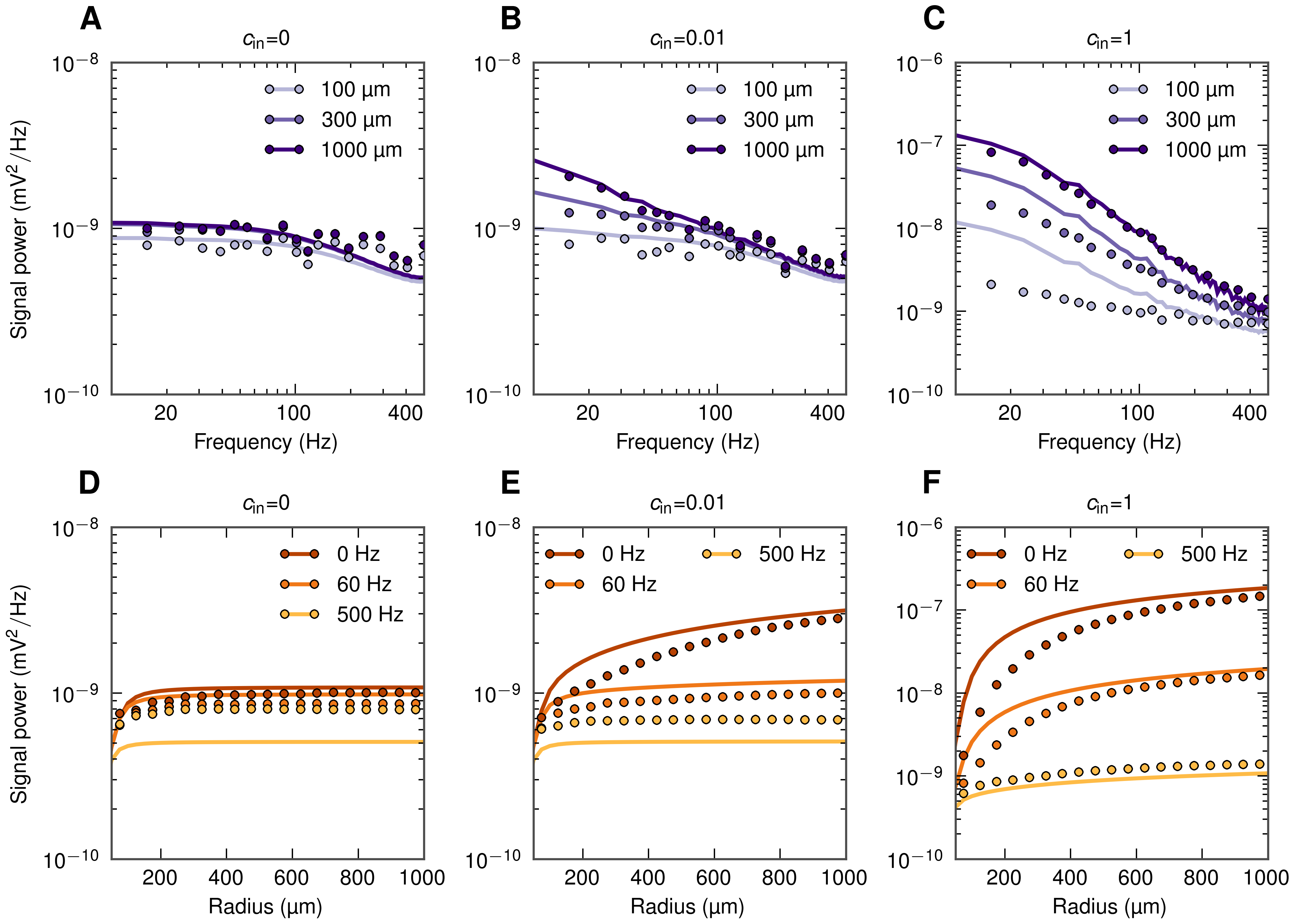}
    \caption{{\bf Power spectral density of population LFP as a function of frequency and the population radius.}
    Full simulation results (dots) and simplified model predictions (lines) for
    the LFP at the center of disc-like populations of layer-5 pyramidal cells receiving basal synaptic input. Three different input correlation levels $c_\text{in}$ are considered. A-C: PSD of population LFP for three population radii $R$. D-F: dependence of power of three different frequency components on the population radius $R$.}
    \label{fig:figure3}
\end{figure}

\subsection{Population LFP} % (fold)
\label{sub:freq_pop_lfp}

As a first step towards exploring the spatial reach of the extracellular potential in our disc-like setup we next show how the population signal emerges from single-cell contributions and investigate frequency-related effects. In Figure~\ref{fig:figure3} we present results both from the full simulation and the simplified model (Equation~\ref{eq:P})
for our example situation with the population of layer-5 cells receiving basal synaptic input.

In Figure~\ref{fig:figure3}A we show the PSD of the LFP produced by differently-sized populations of cells receiving uncorrelated synaptic input. While we observe some low-pass filtering (especially above $\sim$~100~Hz) for all population sizes,
the effect is not particularly strong. Figure~\ref{fig:figure3}D instead shows the PSD for the same uncorrelated situation as a function of the population radius $R$. We observe that the LFP in all frequency bands saturates rather quickly with increasing population size, that is for $R \simeq 100\text{--}200$~$\mu$m. This implies that the contributions from uncorrelated neuronal LFP sources positioned more than a few hundred micrometers away from the electrode are negligible for all frequencies considered.

The situation changes dramatically for the case of correlated synaptic input
(Figure~\ref{fig:figure3}B, \ref{fig:figure3}C, \ref{fig:figure3}E, \ref{fig:figure3}F),
both in terms of amplitude and frequency dependence.
For the case with the maximum input correlations $c_\text{in} = 1$ (Figure~\ref{fig:figure3}C,  \ref{fig:figure3}F), we see that the low-frequency power is up to two orders of magnitude larger than for the corresponding uncorrelated case. Further, a significant  low-pass filtering  effect is seen. For example, the low-frequency power ($\sim$~0~Hz) is an order of magnitude larger than the power at $60$~Hz for $c_\text{in} = 1$ (Figure~\ref{fig:figure3}F). Another observation is that the low-frequency
power grows much faster with increasing population radius than the high-frequency power
(Figure~\ref{fig:figure3}E, \ref{fig:figure3}F). Finally, the power of the population
signal no longer seems to saturate as the population radius increases \cite{Linden:2011ck}.

The predictions from the simplified model agree qualitatively with the full simulation results; however, we observe some clear deviations: First, in Figure~\ref{fig:figure3}D--F we see that the simplified model overestimates the power of the low-frequency components ($\sim$~0~Hz, 60~Hz). This is because the model here uses the approximate power-law shape functions (Equation~\ref{Frast}) which lie above the numerically evaluated shape functions for low frequencies (Figure~\ref{fig:fig_results_intro}C). For high-frequency components (500~Hz), on the other hand, the opposite situation occurs
(results for fitted approximate power-law function not shown).
Second, in case of correlated input the model works better for the larger populations than for smaller ones. This is as expected given the present procedure for calculating the LFP coherence $c_\Phi(f)$ used in the simplified model: here this LFP coherence $c_\Phi(f)$ was extracted from the full population ($R=1000$~$\mu$m) simulations, and the value obtained is not surprisingly a poor approximation when applied to populations which are much smaller. With $c_\Phi(f)$ calculated for each population radius $R$ separately, the simplified model predictions significantly improve (Figure~S1).

% subsection frequency_content_of_lfp (end)

\subsection{Frequency-dependence of spatial reach} % (fold)

We are now ready to analyze the frequency dependence of the spatial reach of extracellular potential. Following \cite{Linden:2011ck} we define the \emph{spatial reach} as the radius of the subpopulation which yields 95\% of the root mean square amplitude in the population center compared to the largest population considered ($R = 1$~mm). With this definition the spatial reach is easily found from the data presented in Figure~\ref{fig:figure3}D, \ref{fig:figure3}E and \ref{fig:figure3}F as the distance at which the amplitude of the LFP reaches 95\% of the maximum value.

\begin{figure}[!ht]
    \centering
        \includegraphics[width=\linewidth]{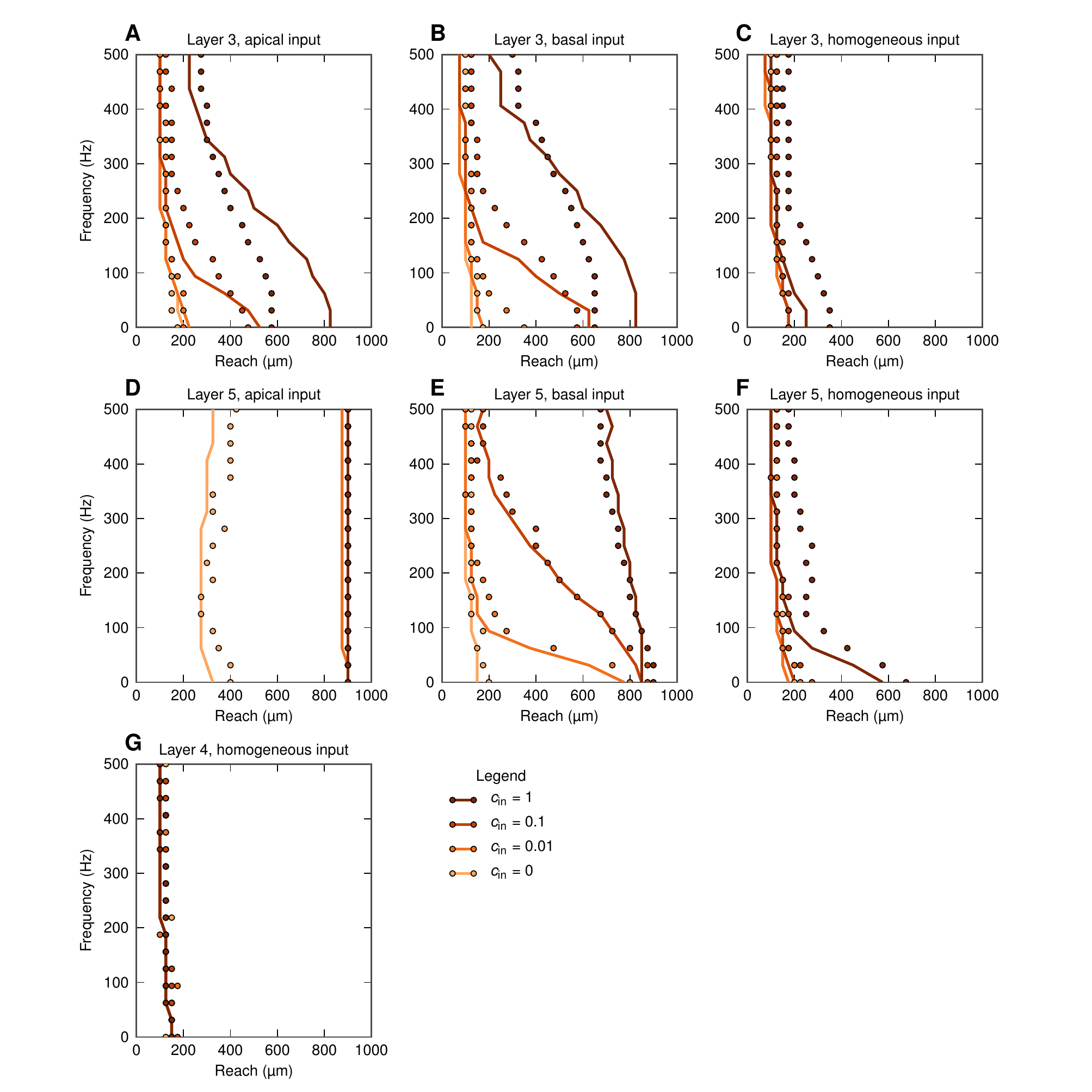}
    \caption{{\bf Spatial reach for different frequency components of LFP.} Spatial reach is defined as the radius of a subpopulation contributing 95\% of the root mean square amplitude of LFP compared to the amplitude for $R=1000$~$\mu$m. LFP is calculated at the soma level at the center of the population. Full simulation results plotted with dots; predictions from the simplified model (Equation~\ref{eq:P}) based on calculated values of $r_\ast$ and $c_{\Phi}$ given in
    Figures~\ref{fig:rstars_popsim_all_long} and \ref{fig:phibars_popsim_all_long}, respectively, are shown with lines. }
    \label{fig:reach_popsim}
\end{figure}

The results for the spatial reach for all seven situations considered are shown in
Figure~\ref{fig:reach_popsim}.
The reach is seen to vary both with the frequency $f$ and the level of input correlation
$c_\text{in}$,
but the specific effects depend sensitively on the cell morphology and synaptic stimulation pattern.
For the pyramidal cells with asymmetric input (either only basal or only apical) the spatial reach grows significantly with increasing input correlations $c_\text{in}$
(Figure~\ref{fig:reach_popsim}A, \ref{fig:reach_popsim}B, \ref{fig:reach_popsim}D, \ref{fig:reach_popsim}E).
The effect is particularly prominent for lower frequencies, i.e., smaller levels of input correlations $c_\text{in}$ are needed to increase the spatial reach significantly.
As a consequence, for certain correlation levels $c_\text{in}$ the spatial reach of the low-frequency components can differ a lot from the spatial reach of the high-frequency components.
For example, in the situation with the layer-5 population receiving basal input with $c_\text{in} = 0.01$, the spatial reach at $100$~Hz is only around 200~$\mu$m, while the low-frequency reach is almost 800~$\mu$m. For the case of homogeneous inputs into pyramidal neurons (Figure~\ref{fig:reach_popsim}C, \ref{fig:reach_popsim}F) these effects are still present, but
seen to be much weaker. For the layer-4 stellate cells the spatial reach is practically independent of the frequency $f$ and the input correlation level $c_\text{in}$, Figure~\ref{fig:reach_popsim}G.

Note that the situation with the layer-5 population receiving only apical input is again somewhat different from the other cases. Here the spatial reach for the uncorrelated input is already quite large ($\simeq 300\text{--}400$~$\mu$m) and the levels of the input correlation required to saturate the spatial reach at a maximum value possible in our setup are significantly smaller.

For the case of uncorrelated input we can obtain analytical expression for the spatial reach from the simplified model. Using
Equations~\ref{eqG0} and ~\ref{Frast} we obtain an explicit formula for $G_0(f,R)$ in terms of the cutoff distance $r_\ast(f)$, the population radius $R$ and $r_\epsilon$. From this, we find in the limit of $r_\epsilon \to 0$, that
 the radius of the subpopulation contributing a fraction $\alpha$ of the asymptotic amplitude ($R \to \infty$) is equal to $r_\ast/\sqrt{3-3\alpha^2}$ (valid for $\alpha^2 > \frac{2}{3}$) . For our choice of $\alpha=0.95$ we find the spatial reach to be $\simeq 1.85 r_\ast$.

% subsection frequency_dependence_of_reach (end)

\subsection{Decay of extracellular potential outside the population} % (fold)

The spatial reach we have discussed above represents an `electrode-centric' point of view: we ask
about the distance from the recording electrode of the neurons setting up the LFP signal.
However, one can also take a `population-centric' approach and instead ask how rapidly
the LFP signal decays with distance outside an active population \cite{Linden:2011ck}.

\begin{figure}[!ht]
    \centering
        \includegraphics[width=\linewidth]{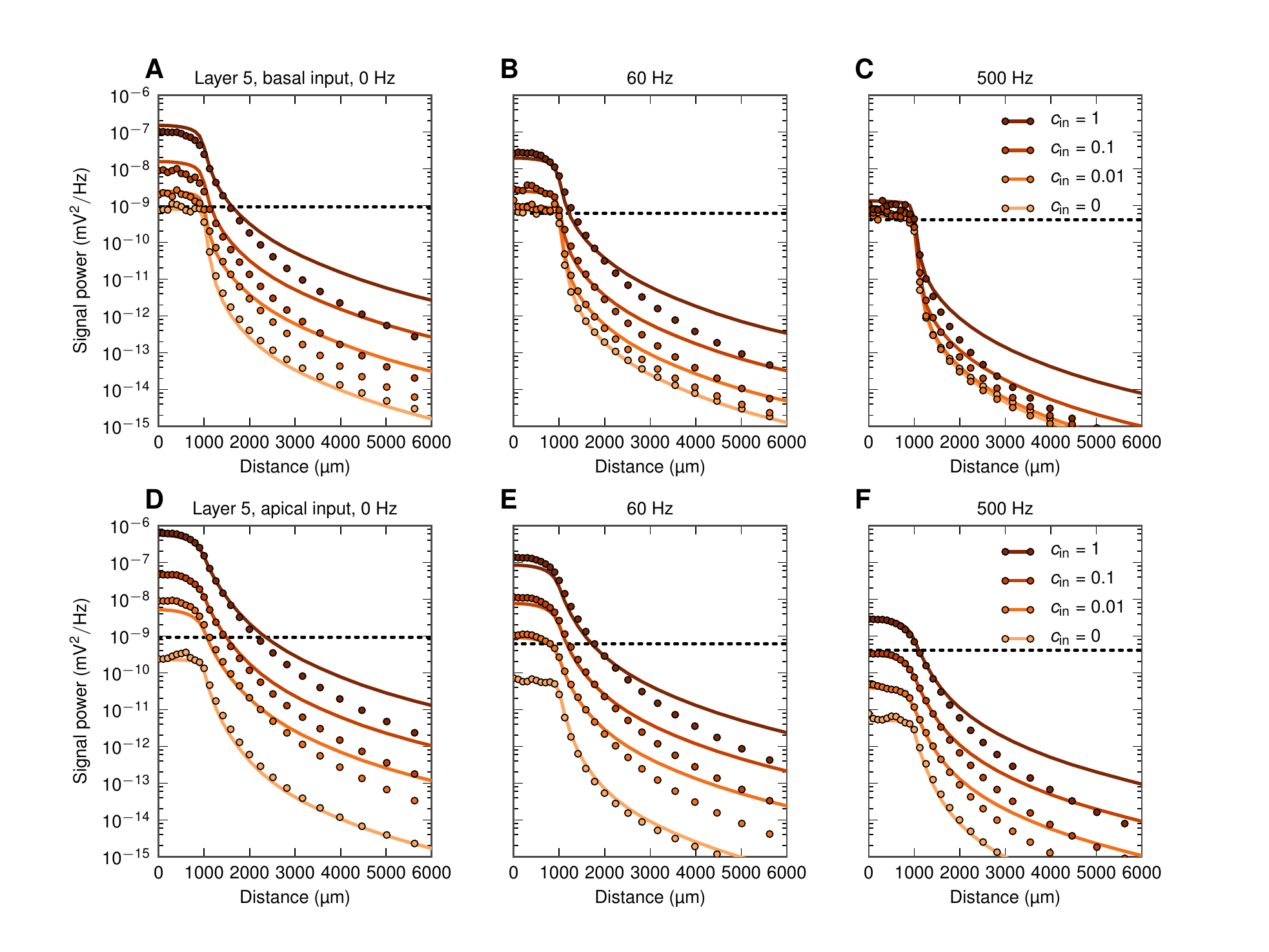}
    \caption{{\bf Decay of extracellular potential outside populations of layer-5 cells with asymmetric input.} Each of the panels shows full simulation results (dots) and predictions from simplified model, Equation~\ref{eq:P} (lines)
    for one frequency band (0, 60, 500~Hz) and four input correlation levels. Horizontal dotted lines indicate `noise level' (power of the signal generated by a population of uncorrelated cells with homogeneous input, see text). A, B, C: basal synaptic input. D, E, F: apical synaptic input.}
    \label{fig:L5_basal_off_decay2}
\end{figure}

In Figure~\ref{fig:L5_basal_off_decay2} we show results for this situation for an example population  of layer-5 cells receiving basal or apical synaptic inputs. The first observation in case of basal synaptic input is that the low- and medium-frequency LFP components ($\sim$~0~Hz, 60~Hz) are significantly boosted, up to two orders of magnitude, by high levels of input correlations $c_\text{in}$ (Figure~\ref{fig:L5_basal_off_decay2}A, \ref{fig:L5_basal_off_decay2}B). This applies both inside and outside of the population. For the high-frequency signal (500~Hz, Figure~\ref{fig:L5_basal_off_decay2}C), however, input correlations are seen to have only a small boosting effect on the signal amplitude.  In the case of apical synaptic inputs the effect of increasing input correlations is seen to be
more uniform across frequency bands, with the high-frequency components (500~Hz) being boosted by roughly the same factor as the low- and medium-frequency LFP components ($\sim$~0~Hz, 60~Hz),
Figure~\ref{fig:L5_basal_off_decay2}D--\ref{fig:L5_basal_off_decay2}F.

The strong boosting of the LFP signal seen for correlated synaptic input for $\sim$~0~Hz (Figure~\ref{fig:L5_basal_off_decay2}A) and
60~Hz (Figure~\ref{fig:L5_basal_off_decay2}B) has direct implications for how recorded LFP signals should be interpreted. As observed in these panels, highly-correlated populations some distance away from the electrode may easily dominate contributions from uncorrelated populations surrounding the electrode. For example, in Figure~\ref{fig:L5_basal_off_decay2}A we observe that the LFP signal
500~$\mu$m \emph{outside} a correlated population with $c_\text{in}$=0.1 is still larger than the contribution recorded \emph{inside} the
same population receiving uncorrelated synaptic inputs ($c_\text{in}$=0). For 60~Hz (Figure~\ref{fig:L5_basal_off_decay2}B) the boosting effect is smaller, but still the signal recorded outside a correlated population may be larger than what is recorded inside an identical population receiving uncorrelated input. The same effect is seen to be even more pronounced for the apical-input case in the lower panels
(Figure~\ref{fig:L5_basal_off_decay2}D--\ref{fig:L5_basal_off_decay2}F), further highlighting that the interpretation of the recorded LFPs in terms of activity in the neurons immediately surrounding the electrode has to be done with caution.

\begin{figure}[!ht]
    \centering
        \includegraphics[width=\linewidth]{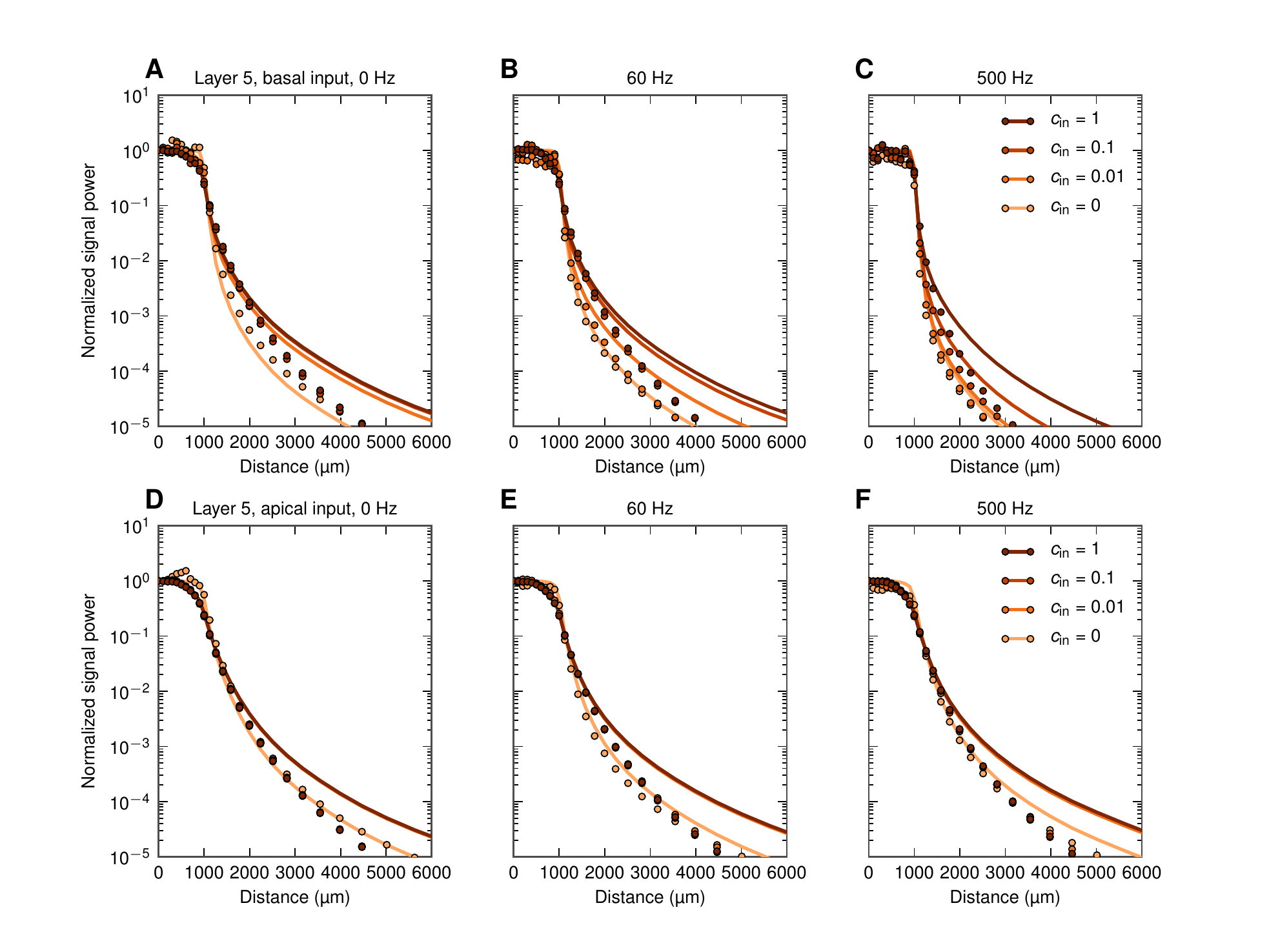}
    \caption{{\bf Decay of extracellular potential outside populations of layer-5 cells with asymmetric input.} Same as Figure~\ref{fig:L5_basal_off_decay2}, but with PSDs normalized to 1 at
    the population center, and the distance axis zoomed in to highlight the behavior around the edge of the population. }
    \label{fig:L5_decay_norm}
\end{figure}

In Figure~\ref{fig:L5_decay_norm} we show the same PSDs as in Figure~\ref{fig:L5_basal_off_decay2}, but normalized to unity at the population center. This illustrates that the decay of the LFP is more abrupt around the population edge in the uncorrelated case than in correlated cases (this is especially prominent for the low-frequency components $\sim$~0~Hz, 60~Hz). This is consistent with an observation made in \cite{Einevoll:2013wq} (see Figure~3.9 therein), namely that in the large-population limit the LFP signal power at the population edge will be reduced to half of power at the center for uncorrelated populations, while it will be reduced to a quarter of the center power for fully correlated populations. Here this difference between the correlated and uncorrelated cases is more pronounced for the low-frequency components, where the coherence $c_\Phi$ is largest.

\begin{figure}[!ht]
    \centering
        \includegraphics[width=\linewidth]{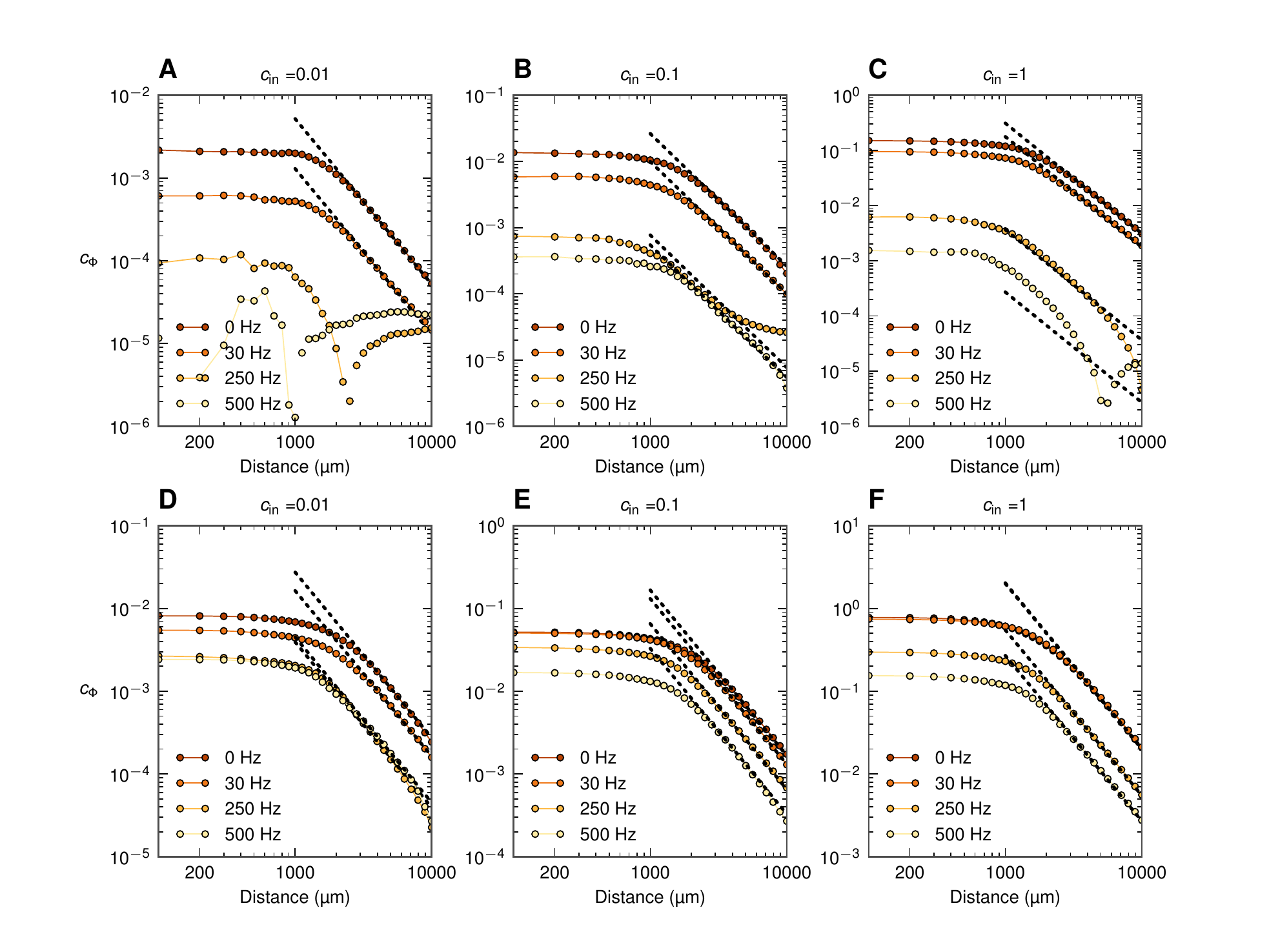}
    \caption{{\bf LFP signal power as functions of frequency and distance from basally-activated pyramidal-cell populations.}
    Colormaps (A, C, E, G) show
    the power of extracellular signal of a population of layer-5 cells receiving basal synaptic input for four levels of input correlation
    $c_\text{in}$ as functions of frequency and distance from center of populations. Black solid and dotted lines denote signal to noise ratio of 0.5 and 0.1, respectively.
    B, D, F, H: power spectra of extracellular signal at different distances, lines: prediction from simplified model in Equation~\ref{eq:P},
    dots: full simulation.
    Thin vertical dotted lines with dots in A, C, E, G denote the distances at which the power spectra are shown, that is, at the center
    (0~$\mu$m), population edge (1000~$\mu$m), and two distances outside ($\sim$~1600~$\mu$m and $\sim$~2500~$\mu$m). }
    \label{fig:cin_cmap}
\end{figure}

We next investigated the related question of detectability, i.e., how far away from a synaptically activated population the generated LFP still can be detected above the ambient LFP `noise'. This noise level will naturally vary between experimental situations, but here we assumed it to be given by the background LFP signal from neurons receiving the same number and type of synaptic inputs, except that (1) the inputs are uncorrelated and (2) homogeneously spread over the neuronal membrane. (The power of this background LFP signal is plotted as dotted lines in Figure~\ref{fig:L5_basal_off_decay2}.)
The frequency-dependent signal decay and detectability outside basally-activated populations are illustrated in the 2D color plots in  Figure~\ref{fig:cin_cmap}. As in Figure~\ref{fig:L5_basal_off_decay2}, the population radius is fixed at $R$=1000~$\mu$m, and we plot the PSD both
inside and outside the population. The lines mark where the signal-to-noise ratio falls below 0.5 (solid line) and 0.1 (dotted line), respectively. Here the signal-to-noise ratio is defined as the ratio between the root mean square amplitudes of the LFP signal (from the basally-activated population) and the LFP noise (from the background population).

A first observation is that for uncorrelated synaptic inputs ($c_\text{in}$=0,~Figure~\ref{fig:cin_cmap}A--\ref{fig:cin_cmap}B), there is very little variation with frequency. Also the detectability of the LFP outside the active population is poor: the signal-to-noise ratio falls to 0.5 about 100~$\mu$m outside the population, and below 0.1 less than 500~$\mu$m outside. The situation is seen to be very different when the populations receive correlated synaptic inputs. Focusing first on the case with the largest level of input correlations ($c_\text{in}$=1, Figure~\ref{fig:cin_cmap}G, \ref{fig:cin_cmap}H), we see that the lower frequencies of LFP extend further outside the population than the higher frequencies. For example, for the near-DC component ($\sim$~0~Hz) the signal-to-noise ratio is seen to be almost 0.5 at a distance of 2000~$\mu$m, i.e., 1000~$\mu$m outside the population edge, and 0.1 as far way as 2000~$\mu$m outside this edge. For the 125-Hz component, on the other hand, the signal-to-noise ratio is reduced to 0.5 as little as 200~$\mu$m outside the population.
The results for the intermediate cases ($c_\text{in}$=0.01, $c_\text{in}$=0.1) depicted in Figures~\ref{fig:cin_cmap}C--\ref{fig:cin_cmap}F are seen to bridge these uncorrelated and strongly correlated cases.

\begin{figure}[!ht]
    \centering
        \includegraphics[width=\linewidth]{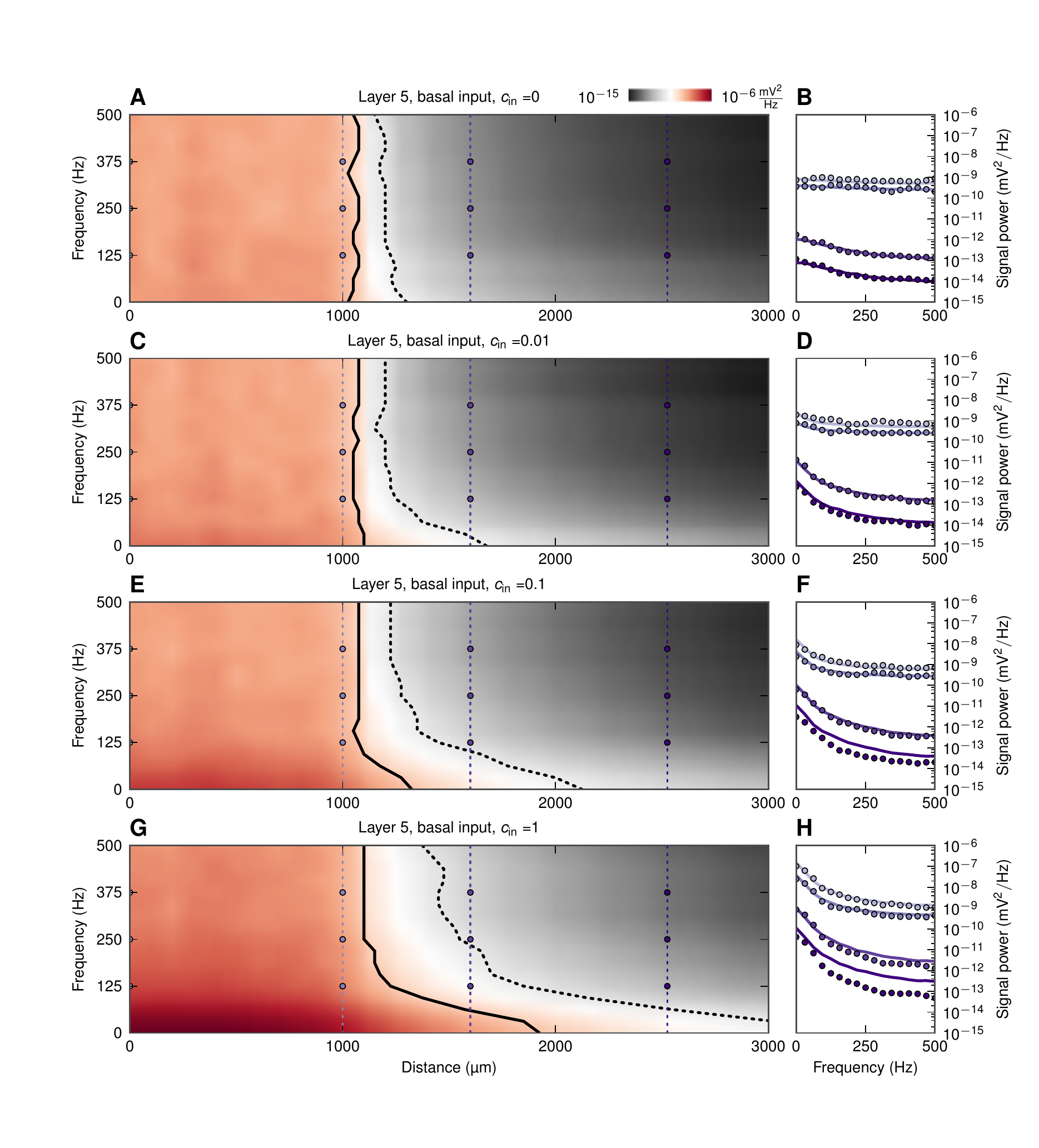}
    \caption{{\bf LFP signal power as functions of frequency and distance from apically-activated pyramidal-cell populations.} Same as Figure~\ref{fig:cin_cmap}, but for a population of layer-5 cells with \emph{apical} synaptic input. }
    \label{fig:cin_cmap2}
\end{figure}

The results for the basally-driven pyramidal cell population in Figure~\ref{fig:cin_cmap} demonstrate a main result from this study,
namely that correlations in synaptic inputs may significantly enhance the amplitude and thus also the detectability of the low-frequency LFP components relative to the high-frequency LFP components. The same effect is observed for the same population when the synaptic inputs are placed solely on the apical part of the neurons, cf. Figure~\ref{fig:cin_cmap2}. However, here a sizable low-pass filtering effect in  detectability is observed also for the case with uncorrelated input (Figure~\ref{fig:cin_cmap2}A, \ref{fig:cin_cmap2}B) due to the intrinsic dendritic filtering effect~\cite{Pettersen:2008p1158, Linden:2010p1109}.
It is also worth noting that populations of layer-5 cells stimulated apically yielded the farthest-reaching LFP signal of all cases analyzed. Note also that the low-pass filtering effect in the boosting of LFP signal with increasing correlations was seen to be largely absent in the case of a spatially homogeneous distributions of synaptic inputs onto populations made of any of our three example neuronal morphologies
(results not shown).

Finally, inspection of Figure~\ref{fig:L5_basal_off_decay2} (and the PSD line plots in Figures~\ref{fig:cin_cmap}~and~\ref{fig:cin_cmap2})
reveals that the predictions from the simplified model (Equation~\ref{eq:P}) agree excellently with the full numerical simulations for the case of uncorrelated input. However, the simplified model systematically overestimates the signal power for correlated populations for positions far outside the active populations. This is because the simplified model predicts a fall-off of the LFP amplitude proportional to $r^{-2}$ in the far-field limit, while in the full simulations the total LFP signal will be dominated by correlated dipoles oriented vertically. As a consequence the functional form of the lateral signal decay will be closer to $r^{-3}$ \cite{Linden:2011ck}.

\begin{figure}[!ht]
    \centering
        \includegraphics[width=\linewidth]{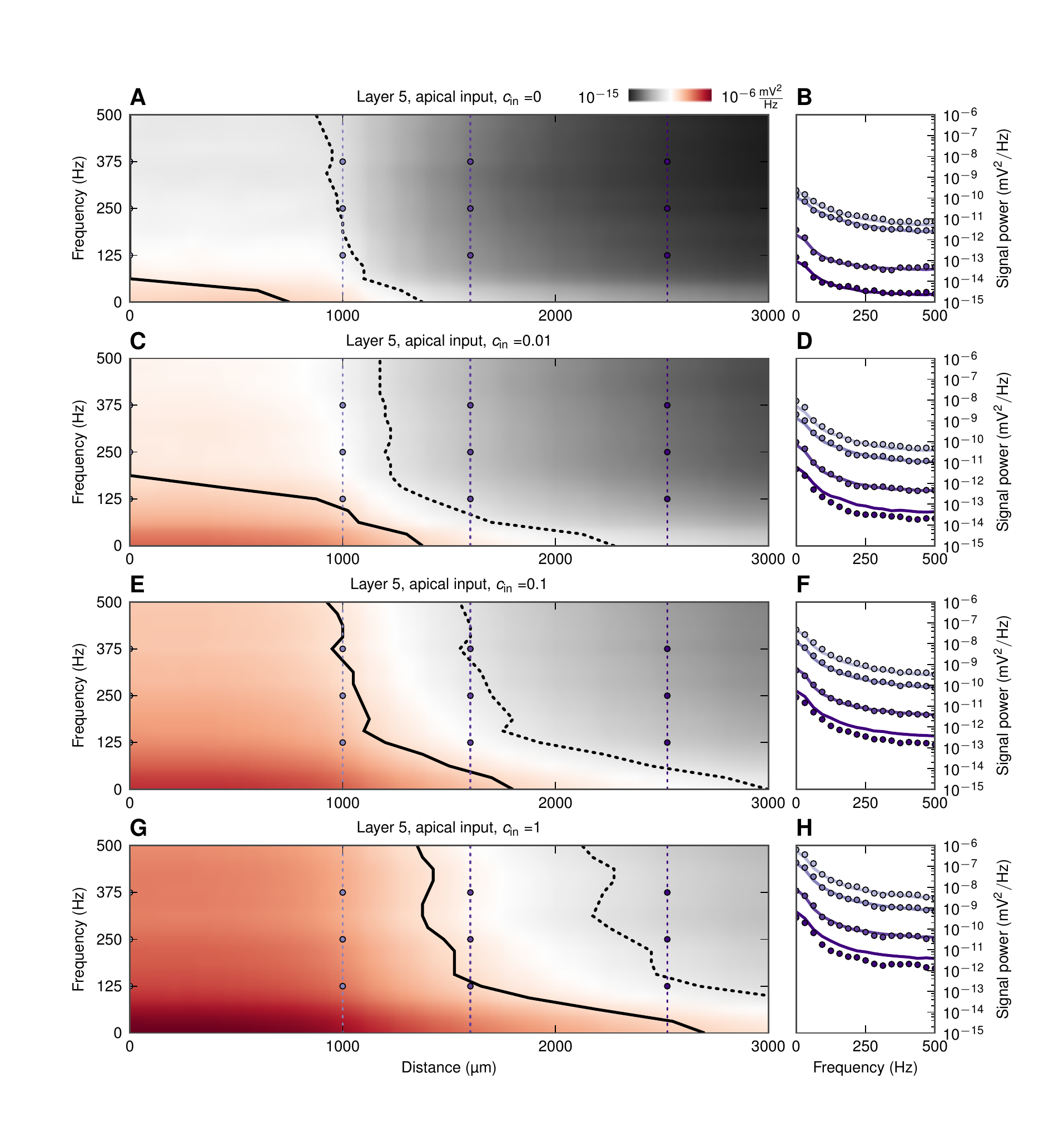}
    \caption{{\bf Population-averaged LFP coherence $c_\Phi$ as a function of distance $X$ from center of population
     of layer-5 pyramidal cells.} A, B, C: basal synaptic input, D, E, F: apical synaptic input. Dots not connected with lines indicate that $|c_{\Phi}|$ is plotted in place of spurious negative values (see Methods). Dashed lines mark $1/X^2$ decay.}
    \label{fig:cphiX}
\end{figure}

This limitation of the simplified model can be remedied by incorporating the fact that the evaluated
population-averaged coherence $c_\Phi(f)$ not only depends on the size of the population $R$ considered, but also on
the electrode position $X$ along the horizontal axis from where it is evaluated, i.e., $c_\Phi(f)=c_\Phi(f;R,X)$.
So far the population-averaged LFP coherence has been evaluated at the population center, i.e., at $X=0$, but when
Equation~\ref{eq:cPhi} is evaluated at other positions $X$, as shown in Figure~\ref{fig:cphiX},
$c_\Phi$ is observed to decay as $1/X^2$ for $X \gg R$. In the formula for the simplified model
in Equation~\ref{eq:P} the power $P$ is seen to be proportional to $c_\Phi G_1$ in the correlation-dominated regime. A modified simplified theory including not only the $X$-dependence of $G_1$ \cite{Linden:2011ck,Einevoll:2013wq}, but also the observed $X$-dependence of $c_\Phi$, indeed predicts the correct far-field $X$-dependence outside the active population (see Figure S8).
The physical interpretation is that the dominance of the LFP signal of the correlated vertical dipoles will be incorporated in the
population-averaged LFP coherence $c_\Phi$.

% section results (end)

\section{Discussion}

In this computational study we have investigated the frequency dependence of the signal power and `locality' of cortical
local field potentials (LFP). While some low-pass filtering effects of the LFP are seen also for populations of cells receiving uncorrelated synaptic inputs or homogeneously distributed correlated synaptic inputs, the large frequency-dependent effects are seen when populations of pyramidal neurons receive correlated and spatially asymmetric inputs (i.e., either only basal or apical). For example, for the case with a layer-5 population receiving correlated, Poissonian synaptic currents (with a white-noise, i.e., flat band, power spectra) onto their basal dendrites, the power of the low-frequency LFP ($\sim$~0~Hz) was seen to be an order of magnitude larger than the corresponding LFP power at 60~Hz. Correspondingly, the low-frequency LFP components was seen to extend much further outside the active population than high-frequency components.

The findings from our comprehensive biophysical simulations using reconstructed neuronal morphologies were backed up
by a simplified model, adapted from \cite{Linden:2011ck}, for generation of population LFP. This model is based on three factors:
(1) the decay of the single-neuron contribution with the distance from the electrode represented by the frequency-dependent
\emph{shape function} $F(f,r)$, (2) the population geometry and density of neuronal LFP sources, and (3) the frequency-dependent \emph{correlation} (or, more precisely \emph{coherence} $c_\Phi(f)$) of the single-neuron LFP contributions from individual neural sources.
Our simple model for the population LFP (Equation~\ref{eq:P}) was found to give quantitatively accurate predictions, implying that it captures the salient features. While some of the observed low-pass filtering could be traced back to single-neuron properties and the
intrinsic dendritic filtering effect \cite{Pettersen:2008p1158, Linden:2010p1109} accounted for by the \emph{shape function} $F(f,r)$, most of the observed low-pass filtering was due to strong low-pass filtering in the \emph{coherence} $c_\Phi(f)$ between the single-neuron LFP contributions: synaptic-input correlations translated into correlated single-neuron LFP contribution to a much larger extent for lower frequencies than for higher frequencies. As a  direct consequence, the low-frequency components of the extracellular potential are significantly boosted in populations with correlated synaptic input. In our model this happens purely because of dendritic filtering, as the synaptic input currents themselves have been tailored to have a flat (white-noise) PSD. With a colored (frequency-dependent) spectrum of the synaptic input, the power spectrum of the LFP would be given as the product of the PSD of this synaptic filter and the PSD from the dendritic filtering investigated here.

A key qualitative finding in our study is that the size of the signal-generating region, i.e.,
the \emph{spatial reach}, may in the case of correlated synaptic input vary strongly with frequency. For the example population in Figure~\ref{fig:sketch} we see that for $c_\text{in}=0.01$, a plausible correlation level in cortical spiking networks (see, e.g., Figure~6 in \cite{Linden:2011ck}), the LFP spatial reach may be reduced from close to the size of the population ($\sim$~800~$\mu$m) for $\sim$~0~Hz to $\sim$~400~$\mu$m for 60~Hz. For uncorrelated input, however, the spatial reach will generally always be small ($\lesssim$ 200~$\mu$m) for all frequencies, with the exception of the case with apical input on large pyramidal cells~(Figure~\ref{fig:reach_popsim}).
Note that in the present simulation scheme the spatial reach is by definition less than 1000~$\mu$m, the size of our model population. Unlike for uncorrelated populations, the LFP power will for correlated populations keep on increasing when the population grows beyond 1000~$\mu$m~\cite{Linden:2011ck}. The present definition of spatial reach (95~\% of the amplitude for $R$=1000~$\mu$m) thus underestimates the true size of the signal-generating region in this case.

In a recent experimental study from macaque auditory cortex \cite{Kajikawa:2011bl} it was observed that different frequency bands spread equally far from a source (cf. Figures~5 and~6 there). There are, however, notable differences between this study and our present approach, making it difficult to compare the results. First, here we investigate the spread of the LFP along cortical layers at the soma level, while in \cite{Kajikawa:2011bl} the spread in vertical direction was studied. Second, and likely more importantly, in \cite{Kajikawa:2011bl} the LFP amplitude at a given latency after stimulation was used to extract LFP decay profiles. In contrast, we here use noise input and consider the root mean square amplitude of LFP over a relatively long time period. Further, the correlation level of the synaptic input, found here to be a critical parameter in determining the frequency dependence, is not known in the situation in \cite{Kajikawa:2011bl}. It is thus difficult to assess whether our results are in accordance, or not.

Our results have direct consequences for the interpretation of observed cross-correlations between extracellular potentials
recorded at different electrodes~\cite{Destexhe:1999,Nauhaus:2009,Ray:2011,Nauhaus:2012,Leopold:2003va,Maier:2010cy}. 
As demonstrated here
the low-frequency LFP signal generated by a population of neurons around one electrode receiving asymmetric synaptic input, may extend a millimeter or more outside the active population (see, e.g., Figure~\ref{fig:cin_cmap}G). Thus measured correlations in the low-frequency LFP components between two electrodes positioned, say, one millimeter apart, may be due to volume conduction effects. However, cross-correlation induced by such volume conduction will, as demonstrated here, have a diminishing spatial range with increasing LFP frequencies. Note also that the magnitudes of the LFP amplitude at the two adjacent electrodes will aid in the interpretation: while volume conduction may propagate the LFP a millimeter or more, the amplitude will rapidly diminish with distance (cf.~Figure~\ref{fig:cin_cmap} and \ref{fig:cin_cmap2}). Thus the observation of large-amplitude LFPs at both electrodes would be an indication that both electrodes are surrounded by strong LFP-generating populations.

In \cite{Freeman:2003vm} the temporal power spectra of the EEG were shown to be well fitted by $1/f^\alpha$ power-law functions with power-law exponents $\alpha$ varying between brain areas: in the frontal lobe $\alpha$ was reported to be $1.78 \pm 0.76$, while in the occipital lobe $\alpha = 1.19 \pm 0.28$. Power laws have also been found in recordings of the LFP, see, e.g., \cite{Milstein:2009hi,Bedard:2006vj}, often with different exponents $\alpha$. In \cite{Bedard:2006vj} $\alpha$ was shown to vary between network states, more specifically between the slow-wave sleep and awake states. In this context it is interesting to note that the PSDs in our Figure~\ref{fig:fig_results_intro}F express approximate power laws with exponents $\alpha$ highly dependent on the degree of coherence. This finding suggests that varying levels of coherence in the synaptic input may be a mechanism underlying the different experimentally observed power laws. This would also be in agreement with the experimental observations that network states with a presumably large coherence (e.g., slow wave sleep in \cite{Bedard:2006vj}) typically express a larger value of $\alpha$ than network states for which the coherence is lower (e.g., awake state in  \cite{Bedard:2006vj}).

In our modeling we have assumed the extracellular medium to have a frequency-independent conductivity, an assumption supported by a recent thorough experimental study of the electrical properties of monkey cortical tissue~\cite{Logothetis:2007kh}. However, if for example low-frequency filtering $\sigma_\text{cond}=\sigma_\text{cond}(f)$ of the extracellular medium should be found~\cite{Bedard:2004iv}, this filtering would superimpose directly on the filtering seen here, i.e., the total LFP filter would be the product of the LFP filter calculated here and the filter from the extracellular medium ($\propto~1/\sigma_\text{cond}(f)$).

Here we have focused on the spatial and spectral properties of LFP signals
triggered by presynaptic spikes that could originate from within the same cortical population or come from other distant brain regions. While not addressed here, it may be that the LFP signal itself influences the timing of these locally generated spikes through ephaptic coupling \cite{Frohlich:2010kt,Anastassiou:2011kp}. That would in turn influence the correlation structure of incoming spikes and thereby also the generated LFP signal. Since our simulations show that both the LFP amplitude and spatial reach is larger for low than for high frequencies, this suggests that if ephaptic effects play a role in cortical processing, they would likely be larger for low than for high frequencies.

In the present analysis we have modeled the dendrites as simple RC-circuits which, in combination with the use of current- based synapses, made the system linear. This greatly facilitated the present frequency-resolved analysis in that the LFPs at different frequencies were effectively decoupled, cf. the standard theory for Fourier analysis of linear systems. The present results also serve as a starting point for the exploration of non-linear effects, for example due to active membrane conductances. Close to the resting potential of the neuron, the active conductances can be linearized, and the neuron dynamics can be described by linear theory with quasi-active membrane modeled by a combination of resistors,
capacitors and inductors (see, e.g., Ch.~10 in \cite{Koch:1999}, Ch.~9 in \cite{Gabbiani:2010}, or \cite{Remme:2011}). At present it is not known to what extent such `generalized' linear schemes will be able to account the LFP generation in real neurons, but the present forward-modeling scheme, applicable for passive and active conductances alike, can be used to explore this question systematically.

% You may title this section 'Methods' or 'Models
% 'Models' is not a valid title for PLoS ONE authors. However, PLoS ONE
% authors may use 'Analysis'
\section{Methods}
\subsection{LFP simulations} % (fold)

The setup of the LFP simulations is almost identical to the scheme used to model cortical population LFPs in \cite{Linden:2011ck}. The main difference is that here we use a much smaller synaptic time constant to achieve an effectively white (flat) power spectrum for the synaptic currents for the frequencies of interest here (less than 500~Hz). We therefore also use a smaller numerical time step. The model parameters are presented in detail (in the format described in \cite{Nordlie:2009gy})
in Tables S1, S2 and S3.  For the reader's convenience we summarize the essential information below.

\subsubsection{Cell models} % (fold)
We analyze three compartmental cell models: the layer-3 and layer-5 pyramidal cells, and layer-4 stellate cells \cite{Mainen:1996cq}, available from ModelDB \cite{Migliore:2003ba}, accession number 2488. We modified the models by removing active conductances and axon segments. The passive parameters of the cells were the following: specific axial resistance $R_a = 150\,\Omega\cdot$cm,
specific membrane resistance $R_m$ = 30 k$\Omega \cdot$cm, specific membrane capacitance
$C_m = 1.0\,\mu$F/cm.

Each simulated cell was stimulated using 1000 excitatory current-based $\alpha$-function synapses with a time constant $\tau$ = 0.1~ms. The synaptic time constant was short enough to ensure that the spectrum of the input current was flat in the studied range. Each synapse was driven by a homogeneous Poisson spike train with the rate of 5 spikes per second. The spike trains driving one cell were independent. For uncorrelated input into the population also the spike trains belonging to each cell were independent, for correlated input they were drawn (without repetitions for each cell) from a common pool consisting of $1000/c_\text{in}$ spike trains. As a result, in case of correlated input each two cells shared $1000\cdot c_\text{in}$ spike trains on average. Note that even for $c_\text{in} = 1$, when each of the cells is driven by the same spike trains, the spike trains will in general be assigned to different synaptic locations.

We simulated activity of cells for either 10200~ms (single-cell shape functions and
LFP in the population's center) or 1200~ms (LFP at points not in the population's center).
The first 200~ms were discarded to avoid start-up artifacts.
We used a fixed time step of 1/64~ms, and recorded the results of the simulation (transmembrane currents in all compartments) with 1~ms time step (sampling frequency 1~kHz).

For the pyramidal cells we employed three stimulation patterns: the synapses were distributed either in the apical or basal part, or homogeneously throughout the whole dendritic tree (in each case the probability of attaching a synapse in a given compartment was proportional to its surface area). We used the same layer boundaries and soma depths as in \cite{Linden:2011ck}.
% subsubsection cell_models (end)

\subsubsection{Calculation of LFP} % (fold)

The extracellular electric potential was calculated using the line-source method \cite{Holt:1998un,Holt:1999wi}, resulting from integration of Equation~\ref{eq:pointsource} over linear dendritic segments. We assumed a purely resistive, homogeneous, isotropic and infinite extracellular medium, and an ideal point electrode (no filtering), placed at the soma level. In single-cell simulations the electrode was placed at a distance (between 10~$\mu$m and 10000~$\mu$m) from a single cell, in population simulations it was placed either at the center of the population or at 31 points placed between 0~$\mu$m and 10000~$\mu$m from the center. To obtain the model LFP at the center of differently-sized populations we summed contributions from different subsets (cells located closer to the electrode than some distance) of the same full ($R=1000$~$\mu$m) population.

% subsubsection calculation_of_lfp (end)

\subsubsection{Single-cell shape functions} % (fold)

To obtain single-cell shape functions (Figure~\ref{fig:fig_results_intro}A) we calculated the LFP at different distances from a single cell, then calculated power spectra of these signals. The final curves were obtained by averaging power spectra from 100 simulations for each distance.

% subsubsection single_cell_decay_curves (end)

\subsubsection{Population simulations} % (fold)

We simulated populations of $N=10000$ identical neuron. The cells were placed homogeneously within a disc of $1$~mm radius  at the same depth. Each cell was rotated randomly along the vertical axis.

% subsubsection population_simulations (end)

\subsubsection{Software} % (fold)
We performed the simulations using the NEURON simulator (\cite{Carnevale:2006vua}, \path{www.neuron.yale.edu})
and the Python (\path{www.python.org}) interface to NEURON \cite{Hines:2009gh}; we also used NeuroTools (\path{neuralensemble.org/trac/NeuroTools/}). The calculations of extracellular field were performed using LFPy \cite{Linden:2011jg} --- Python package for modeling of LFP.
% subsubsection software (end)

% subsection lfp_simulations (end)

\subsection{Derivation of the mean-field model} % (fold)

To derive the formula in Equation~\ref{eq:P} for the power spectral density (PSD) of the extracellular signal in the center of the population we start we the assumption that the PSD of the contribution of the $i$-th cell at given frequency $f$ ($\Phi_i(f)$) may be factorized as
\begin{equation}
    \label{ass1}
    |\Phi_i(f)|^2 \approx
 \sigma_\xi^2(f) F_i^2(f),
\end{equation}
where
$\sigma_\xi^2$ is the PSD of the input current, and $F_i(f)$ is the frequency-dependent \emph{shape function} of the $i$-th cell.
We also assume that the shape function $F$ depends only on frequency and distance from the center, that is:
\begin{equation}
    \label{ass2}
    F_i(f) = F(f, r_i).
\end{equation}

Let us compute the PSD of the population signal $\Phi(f)$ (dependence on frequency $f$ dropped below for convenience):
\begin{equation}
   P = |\Phi(f)|^2 = \Phi^\ast\Phi = \Big(\sum_{i=1}^N \Phi_i^\ast \Big)\Big(\sum_{j=1}^N \Phi_j \Big) = \sum_{i=1}^N \Phi_i^\ast\Phi_i + \underset{i\neq j}{\sum_{i=1}^N\sum_{j=1}^N} \Phi_i^\ast\Phi_j.
\end{equation}
We now use Equations \ref{ass1} and \ref{ass2} to express $P$ in terms of shape functions and the PSD of the input current, note the trick in the double sum:
\begin{equation}
    P = \sigma_\xi^2\left(\sum_{i=1}^N F(r_i)^2
    + \underset{i\neq j}{\sum_{i=1}^N\sum_{j=1}^N} \frac{\Phi_i^\ast}{|\Phi_i|}\frac{\Phi_j}{|\Phi_j|} F(r_i)F(r_j)\right).
\end{equation}
We further assume that the coherence term $\frac{\Phi_i^\ast}{|\Phi_i|}\frac{\Phi_j}{|\Phi_j|}$ may by replaced by its population average
over $N(N-1)$ pairs, hence we can move it in front of the double sum:
\begin{equation}
    \label{eq:dercoh}
P = \sigma_\xi^2\Bigg(\sum_{i=1}^N F(r_i)^2 +
\underbrace{\frac{1}{N(N-1)}
\underset{i\neq j}{\sum_{i=1}^N\sum_{j=1}^N}
\frac{\Phi_i^\ast}{|\Phi_i|}\frac{\Phi_j}{|\Phi_j|}}_{=:c_\Phi}
\underset{i\neq j}{\sum_{i=1}^N\sum_{j=1}^N}F(r_i)F(r_j)\Bigg).
\end{equation}
As marked in Equation~\ref{eq:dercoh}, we denote the population-averaged coherence by $c_\Phi$.
We further rewrite $P$ as
\begin{equation}
    P= \sigma_\xi^2\Bigg[\sum_{i=1}^N F(r_i)^2 +
    c_\Phi
    \Big(\Big|\sum_{i=1}^N F(r_i)\Big|^2 - \sum_{i=1}^N F(r_i)^2
    \Big)\Bigg],
\end{equation}
 and finally
 \begin{equation}
     P= (1-c_\Phi)\underbrace{\sigma_\xi^2\sum_{i=1}^N F(r_i)^2}_{=:G_0} +
     c_\Phi \underbrace{\sigma_\xi^2
     \Big|\sum_{i=1}^N F(r_i)\Big|^2}_{=:G_1} .
 \end{equation}

If we assume approximate, power-law shape functions $F(f,r)$ parametrized by the cutoff distance $r_\ast(f)$ (Equation~\ref{Frast}), and change sums to integrals as in Equations~\ref{eqG0} and \ref{eqG1} (limit of large number of cells), then the functions $G_0(f, R)$ and $G_1(f,R)$ have the
following closed-form representation \cite{Einevoll:2013wq}:
\begin{eqnarray}
    G_0(f, R) & = & \begin{cases}
        F_0^2\rho\pi R^2
            &\text{ if  }R\le r_\epsilon,\\
        F_0^2\rho\pi r_\epsilon(2R-r_\epsilon)
            &\text{ if  } r_\epsilon < R \le r_\ast,  \\
        F_0^2\rho\pi r_\epsilon\big(3r_\ast-r_\epsilon-r_\ast^3/R^2\big)
            &\text{ if  } r_\ast \le R,
    \end{cases}     \\
    G_1(f, R) & = & \begin{cases}
        F_0^2\rho^2\pi^2 R^4
            &\text{ if  }R\le r_\epsilon,\\
        \frac19 F_0^2\rho^2\pi^2 \big(r_\epsilon^2-4r_\epsilon^{1/2}R^{3/2}\big)^2
            &\text{ if  } r_\epsilon < R \le r_\ast,  \\
        \frac19 F_0^2\rho^2\pi^2 r_\epsilon\Big(r_\epsilon^{3/2}-\big(4+6\log(R/r_\ast)\big)r_\ast^{3/2}\Big)^2
            &\text{ if  } r_\ast \le R,
    \end{cases}
\end{eqnarray}
which we used for calculating predictions from the simplified model.
The model can be modified to calculate the power of the signal outside the center of the population, i.e., at positions offset from the center by the distance $X$. In that case, the function $F(f, R)$ in \eqref{eqG0} and \eqref{eqG1} has to be replaced by $F(f, |r-X|)$. It is no longer easy to obtain closed-form formulae for $G_0$ and $G_1$ in terms of $r_\ast$, and we used the
(non-parametric) shape curves obtained from the simulations, as the final integration had to be done numerically anyway.

% subsection mean_field_model (end)

\subsection{Data analysis} % (fold)

\subsubsection{Population-averaged LFP coherence} % (fold)
\label{sub:mean_lfp_coherence}
It is hard to estimate the population-averaged LFP coherence $c_\Phi$ directly as an average of pairwise coherences between the single-cell contributions to the LFP. Therefore, we used the same technique as in \cite{Linden:2011ck} (Equations 14 and 15 therein), ending up with
\begin{equation}
    \label{eq:cPhi}
    c_\Phi(f) = \frac{\left| \sum_{i=1}^N \frac{\Phi_i(f)}{|\Phi_i(f)|} \right|^2-N}{N(N-1)} .
\end{equation}
Coherence is always positive; however, the population-averaged coherence $c_\Phi$ estimated using Equation~\ref{eq:cPhi} may take spurious negative value (for example because finite-length signals are used). This does not mean that $c_\Phi$ is truly negative, but rather that the value is too small to be estimated reliably from the amount of data available. In such cases we plotted $|c_\Phi|$ in figures.

Note that $\Phi_i(f)$ in Equation~\ref{eq:cPhi} may be evaluated either at the population center, or at a lateral position $X>0$; as a result we will get either $c_\Phi(f; X=0)$ or
$c_\Phi(f; X)$, see the last Section of Results.

% subsubsection mean_lfp_coherence (end)
\subsubsection{Frequency analysis} % (fold)
To calculate the power spectral densities we used the Welch’s average periodogram method (the \path{matplotlib.mlab.psd} function from Matplotlib \cite{citeulike:2878517}). We used a Hanning window of length 32 or 128 time steps (32 or 128~ms) and overlap between blocks equal to the half of the window length, which resulted in 17 (or 65) equally spaced frequency bins between 0 and 500~Hz. When calculating the population-averaged LFP coherence, Equation~\ref{eq:cPhi}, we calculated the discrete Fourier transform and binned the resulting $c_\Phi$ in the same frequency bins as resulting from the Welch's average periodogram method.

% subsubsection frequency_analysis (end)

\subsubsection{Spatial reach of LFP} % (fold)

To obtain the spatial reach of the LFP we calculated the power spectral density $P(f,R)$ of the population LFP as a function of frequency $f$ and population radius $R$ (taking values between $0$ and 1000~$\mu$m in 25~$\mu$m increments). The spatial reach at given frequency was defined as the smallest radius $R^\ast$ for which the amplitude $\sigma_\Phi(f, R^\ast)$ is larger than $95\%$ of the amplitude calculated for the full population.

% subsubsection reach_of_lfp (end)

\subsubsection{Software} % (fold)
Data analysis was performed using NumPy and SciPy Python packages \cite{Oliphant:2007ud} and IPython \cite{Perez:2007wf}. Plotting was done using Matplotlib \cite{citeulike:2878517}.
% subsubsection software (end)

% subsection data_analysis (end)

% section methods (end)

% Do NOT remove this, even if you are not including acknowledgments
\section*{Acknowledgments}
We acknowledge support from the The Research Council of Norway (eVita, NOTUR, Yggdrasil), Polish Ministry of Science and Higher Education (grants
N~N303~542839 and IP2011~030971), and  the EU Grant 269921 (BrainScaleS). The project has been implemented with support granted by Iceland, Liechtenstein, and Norway, through a grant from the funds of the Financial Mechanism of the European Economic Area and the Norwegian Financial Mechanism under the Scholarship and Training Fund.

\cleardoublepage
\setcounter{figure}{0}
\setcounter{table}{0}
\makeatletter 
\renewcommand{\thefigure}{S\@arabic\c@figure} 
\renewcommand{\thetable}{S\@arabic\c@table} 

\textbf{Supplementary Information}

\

\begin{large}
\noindent
\textbf{Frequency dependence of signal power
\\ and spatial reach
of the local field potential}
\end{large}

\

\noindent
Szymon Łęski,
Henrik Lindén,
Tom Tetzlaff,
Klas H. Pettersen,
Gaute T. Einevoll

\vspace{2cm}
% 
% \begin{large}
% \noindent
% \textbf{Inventory of Supplementary Information}
% \end{large}
% 
% \
% 
% \begin{enumerate}
% %
% \item \textbf{Supplementary Figures}:
% \begin{itemize}
% \item \emph{Figure~S1}: Alternate version of Figure 6 from the paper; here the coherence $c_\Phi$ is estimated not just once for the full population, but in a radius-dependent fashion; in effect the simplified model is closer to the full simulations.    
% \item \emph{Figures~S2--S7}: Ingredients of the simplified LFP model for the remaining cell types and stimulation patterns.  A. Spatial decay in lateral direction for the squared single-cell shape functions $|F (f, r)|^2$ for three different frequencies $f = 0, 60$ and $500$ Hz. B. Single-cell LFP spectra $|F (f, r)|^2$ for three different lateral distances from the soma (dotted vertical lines in A). C. Power spectra P (f, R) of the compound LFP (R = 1 mm); dots: simulation; lines: predictions from simplified model, eq. (4)
% These figures correspond to panels A, B, and E of Figure 6 in the paper.  
% \end{itemize}
% %
% \item \textbf{Supplementary Tables}:
% \begin{itemize}
% \item \emph{Tables S1--S3}: Specify details of the numerical simulations. 
% \end{itemize}
% %
% % \item \textbf{Supplementary References}
% \end{enumerate}
% 
% 
% \clearpage

% \section{Supplementary Figures}

\ifpdf
\DeclareGraphicsExtensions{.pdf, .jpg, .tif}
\else
\DeclareGraphicsExtensions{.eps, .jpg}
\fi     

\begin{figure}[htbp]
    \centering
     \includegraphics[width=\textwidth]{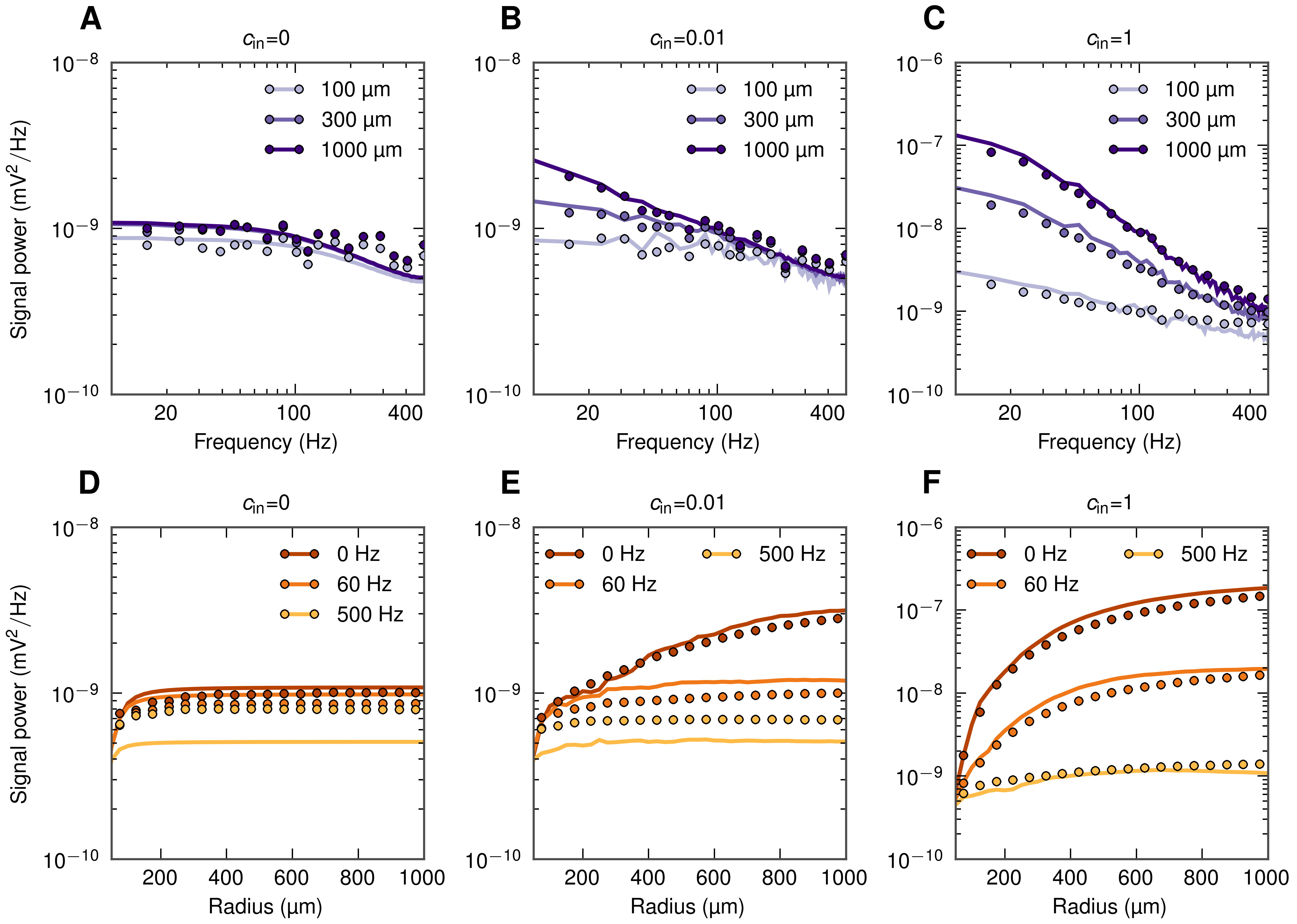}
    \caption{Power spectral density of population LFP as a function of 
    frequency and the population radius. 
    Full simulation results (dots) and simplified model predictions (lines) for
    the LFP at the center of disc-like populations of layer-5 pyramidal cells receiving basal  
    synaptic input. Three different input correlation levels $c_\text{in}$ are considered. A-C: PSD
    of population LFP for three population radii $R$. D-F: dependence of power of three different
    frequency components on the population radius $R$. 
  This is an alternate version of Figure 6 from the paper; here the coherence $c_\Phi$ is estimated
   not just once for the full (R = 1000 $\mu\text{m}$) population, but in a radius-dependent
    fashion, for each population radius $R = 25, 50, 75, \ldots  1000 \mu\text{m}$ separately. In
effect the simplified model predictions are closer to the full simulations than in Figure 6.}
    \label{fig:figure3_supp}
\end{figure}

        \begin{figure}[htbp]
            \centering
                \includegraphics[width=\textwidth]{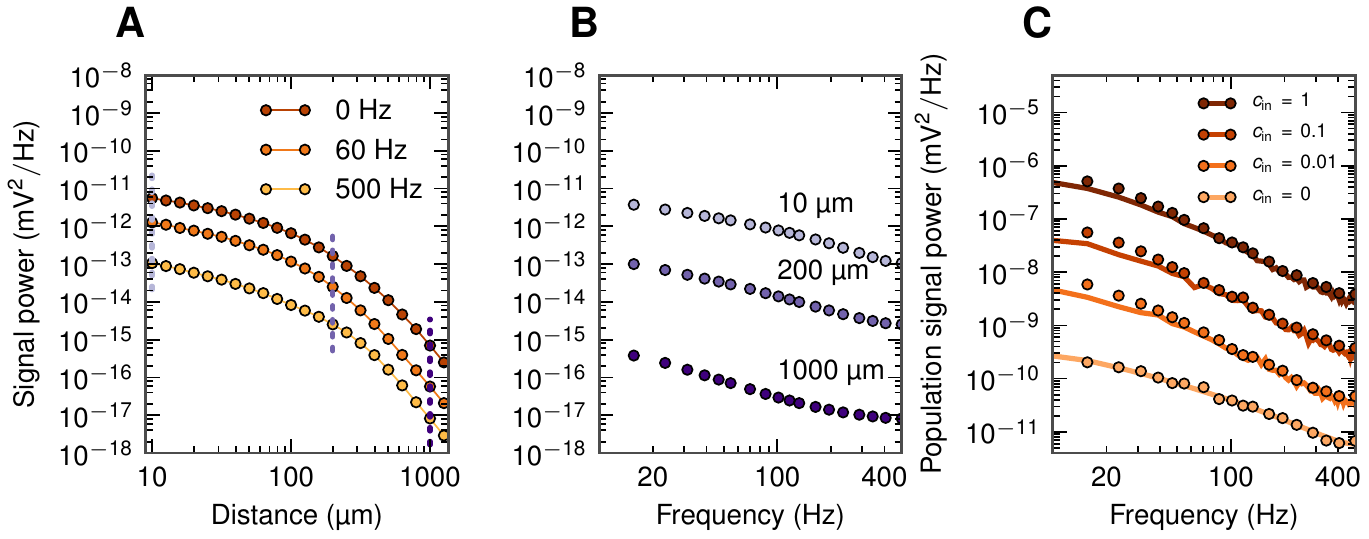}
            \caption{    The shape function $|F(f,r)|^2$ and the population LFP power spectra for  
                layer-5 cells with apical input.                                         
                A. Spatial decay in lateral direction for the squared single-cell shape functions $|F(f, r)|^2$ for three different frequencies f = 0, 60 and 500 Hz. B. Single-cell LFP spectra $|F(f, r)|^2$ for three different lateral distances from the soma (dotted vertical lines in A). C. Power spectra $P(f, R)$ of the compound LFP (R = 1 mm); dots correspond to simulation; lines correspond to predictions from the simplified model.}
            \label{fig:figure2_supp_2}
        \end{figure}
        
 \begin{figure}[htbp]
    \centering
        \includegraphics[width=\textwidth]{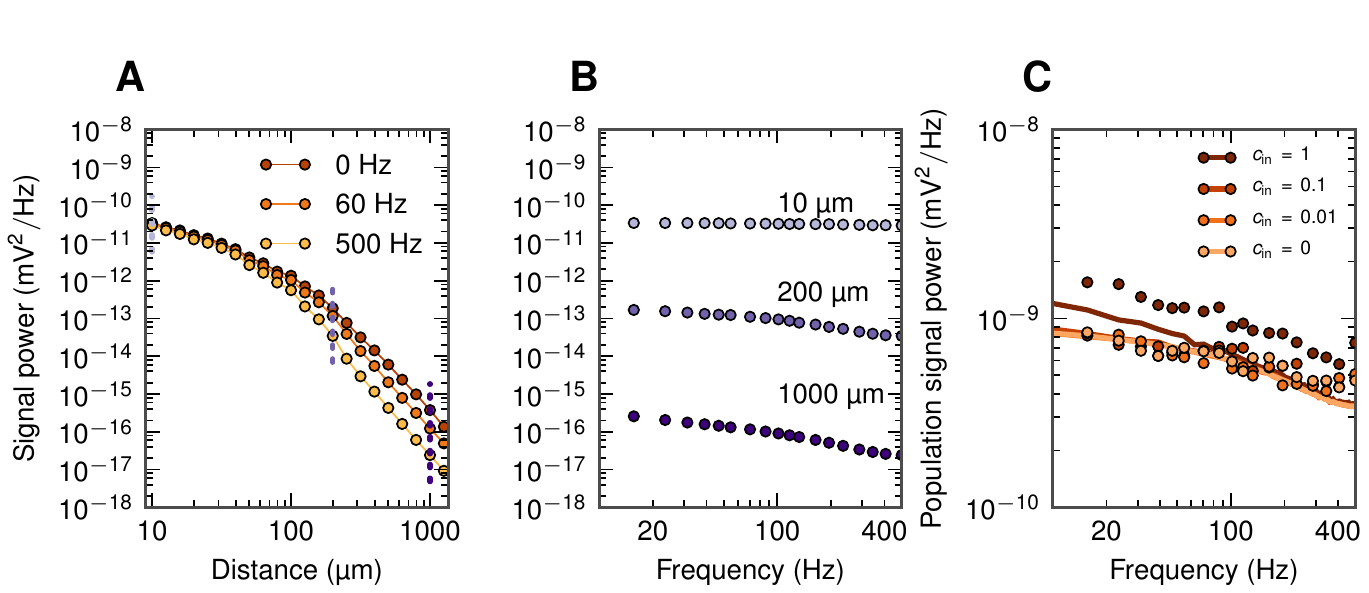}
    \caption{The shape function $|F(f,r)|^2$ and the population LFP power spectra for  
        layer-5 cells with homogeneous input. See Caption of Figure S2.}
    \label{fig:figure2_supp_3}
 \end{figure}
             
\begin{figure}[htbp]
    \centering
        \includegraphics[width=\textwidth]{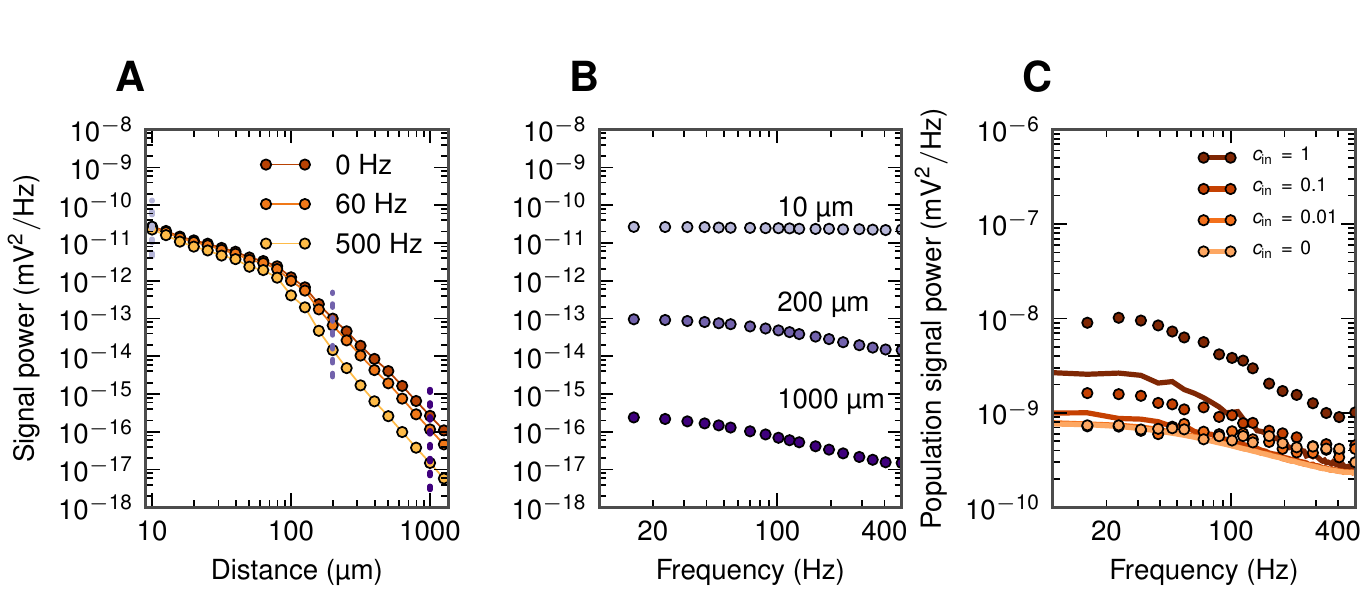}
    \caption{The shape function $|F(f,r)|^2$ and the population LFP power spectra for  
        layer-3 cells with apical input. See Caption of Figure S2.}
    \label{fig:figure2_supp_4}
\end{figure}

\begin{figure}[htbp]
    \centering
        \includegraphics[width=\textwidth]{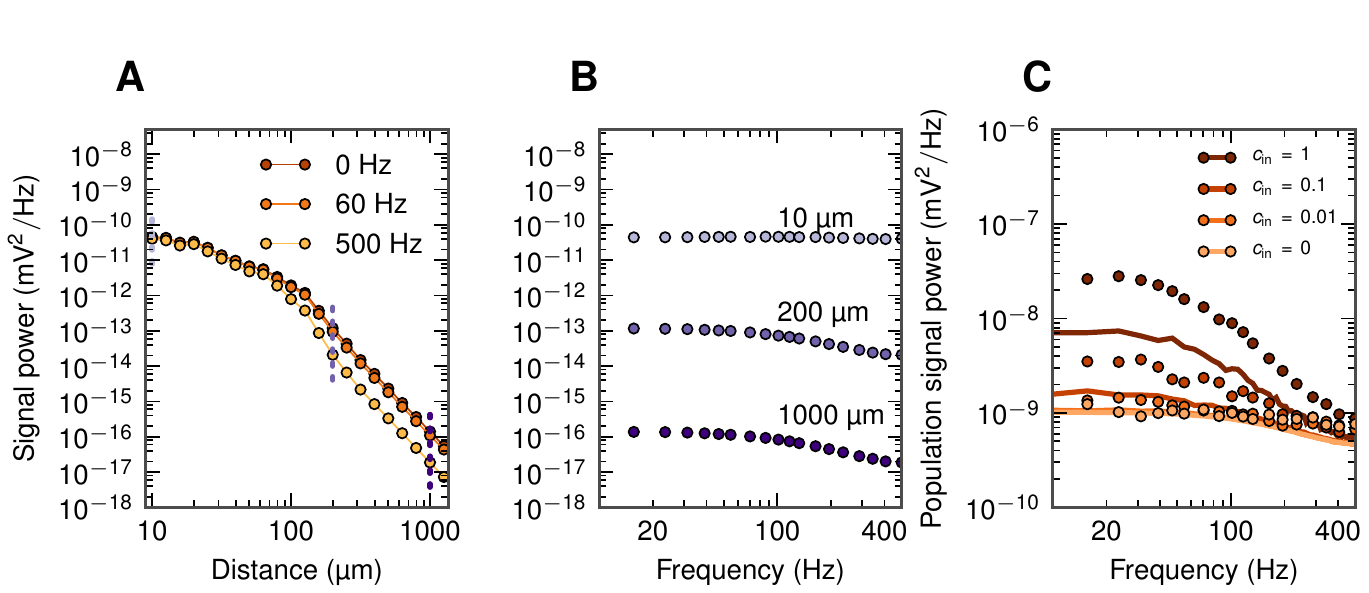}
        \caption{The shape function $|F(f,r)|^2$ and the population LFP power spectra for  
            layer-3 cells with basal input. See Caption of Figure S2.}
    \label{fig:figure2_supp_5}
\end{figure}

\begin{figure}[htbp]
    \centering
        \includegraphics[width=\textwidth]{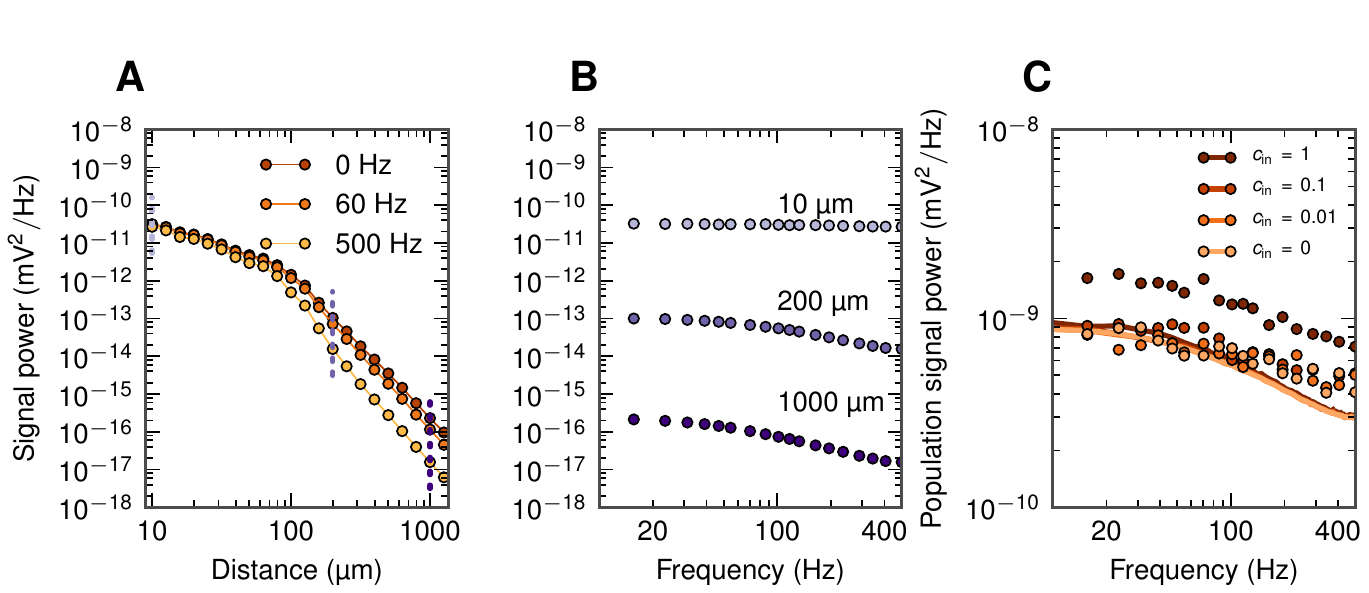}
        \caption{The shape function $|F(f,r)|^2$ and the population LFP power spectra for  
            layer-3 cells with homogeneous input. See Caption of Figure S2.}
    \label{fig:figure2_supp_6}
\end{figure}

\begin{figure}[htbp]
    \centering
        \includegraphics[width=\textwidth]{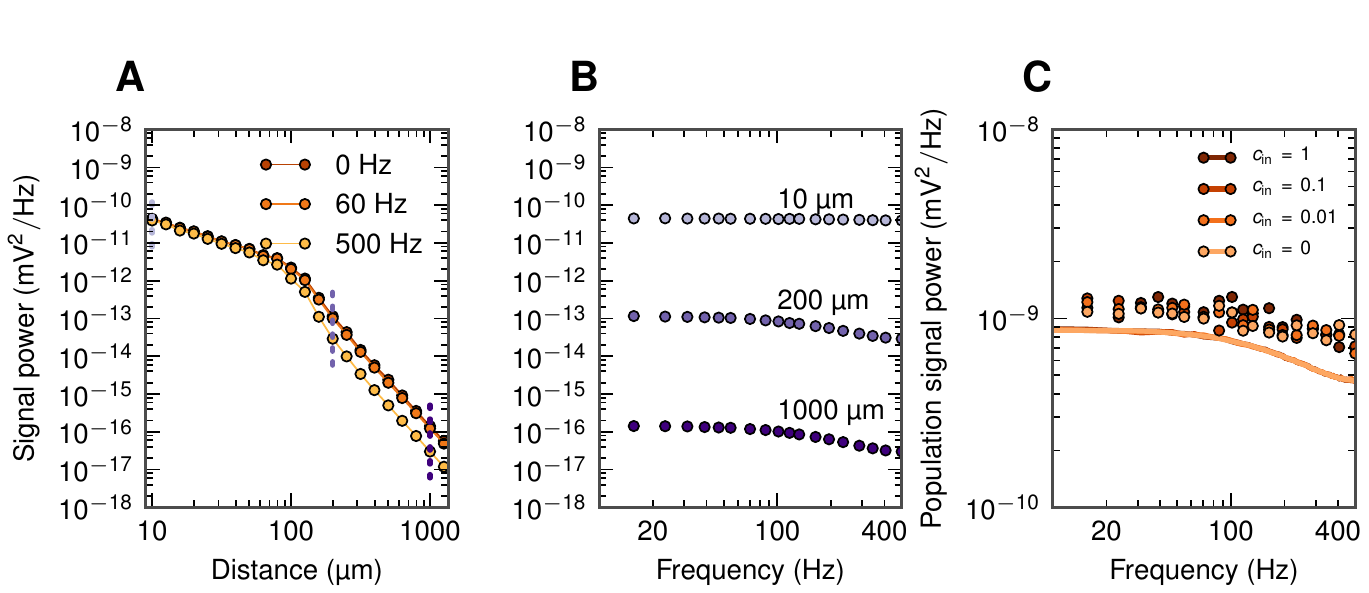}
        \caption{The shape function $|F(f,r)|^2$ and the population LFP power spectra for  
            layer-4 cells with homogeneous input. See Caption of Figure S2.}
    \label{fig:figure2_supp_7}
\end{figure}

\begin{figure}[htbp]
    \centering
        \includegraphics[width=\textwidth]{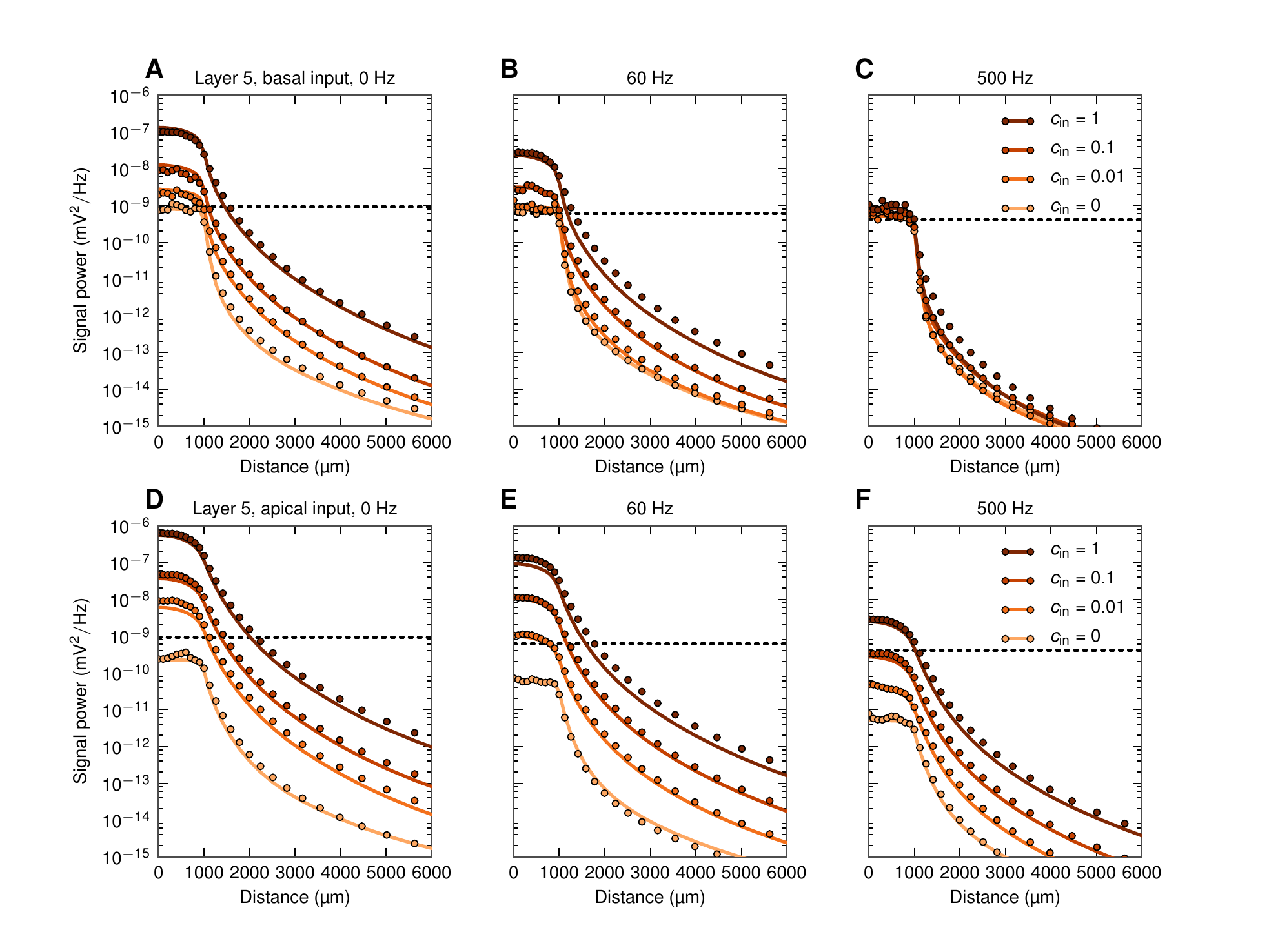}
        \caption{Decay of extracellular potential outside populations of layer-5 cells with asymmetric input. Each of the panels shows full simulation results (dots) and predictions from simplified model Equation 5 (lines) for one frequency band (0, 60,
        500 Hz) and four input correlation levels. Horizontal dotted lines indicate ‘noise level’ (power of the signal generated by a population of uncorrelated cells with homogeneous input, see text). A, B, C: basal synaptic input. D, E, F: apical synaptic input. 
        This is an alternate version of Figure 8 from the paper, here the population-averaged
        coherence $c_\Phi$ depends also on the distance. In effect the simplified model predictions are closer to the full simulations than in Figure 8.}
    \label{fig:figureS8}
\end{figure}

\clearpage

\def\tabspace{0.5ex}
\begin{table}[ht!]

  %% Summary
  \begin{tabularx}{\linewidth}{|p{0.2\linewidth}|X|}
    \hline\modelhdr{2}{A}{Model summary}\\\hline
    \bf Population& cylindrical homogeneous cortical populations \\\hline
    \bf Neuron & passive multi-compartment neuron models \\\hline
    \bf Synapse & current based, alpha-shaped postsynaptic current with short time constant\\\hline
    \bf Input & uncorrelated/correlated Poisson spike train input or spike train input generated by laminar network model \\\hline
    \bf Measurements & simulated LFP\\\hline
  \end{tabularx}\vspace{\tabspace}

  %% Population
  \begin{tabularx}{\linewidth}{|p{0.2\linewidth}|X|}
    \hline\modelhdr{2}{B}{Population}\\\hline
    \bf Type & separate homogeneous populations consisting of $N$ neurons  \\
    \bf & \emph{Population types}: L3 pyramidal cell population, L4 stellate cell population, L5 pyramidal cell population\\\hline
    \bf Geometry & cylinder of radius $R$ (`cortical column') subdivided into layers (see Table~S3A) \\\hline
    \bf Cell positions & - random soma positions on a disc at soma depth $z_k$ in vertical midpoint of corresponding cell type $k$, soma density $\rho = N/(\pi R^2)$ \\
    \bf & - random cell rotations along vertical cylindrical axis\\\hline
    \bf Parameters&   $N$, $R$, $z_k$, layer boundaries \\\hline

  \end{tabularx}\vspace{\tabspace}

  %% Neuron
  \begin{tabularx}{\linewidth}{|p{0.2\linewidth}|X|}
    \hline\modelhdr{2}{C}{Neuron}\\\hline
    \bf Type &  passive multi-compartment neuron models with reconstructed morphologies\\\hline
    \bf     Morphology & - L3 pyramidal cell \\
               & - L4 stellate cell \\
               & - L5 pyramidal cell \\
               & from [63], %\cite{Mainen:1996p622}, 
               downloaded from ModelDB, accession number 2488 \\
               & axon compartments were removed \\\hline

    \bf Neuron \newline dynamics & non-spiking neurons with passive membrane with specific membrane resistance $R_\text{m}$, specific axial resistance $R_\text{a}$, and specific membrane capacitance $C_\text{m}$ \\\hline

    \bf Compartments & segments length shorter than one tenth of electrotonic length for 100 Hz resulting in 549 compartments for the L3 cell, 343 compartments for the L4 cell and 1072 compartments for the L5 cell (for chosen passive parameters and morphologies,
    see Table S3) \\\hline
    \bf Parameters & $R_\text{m}$, $R_\text{a}$, $C_\text{m}$\\\hline

  \end{tabularx}\vspace{\tabspace}

\caption{Summary of population model used in LFP simulations. Continues in Table \prettyref{table:LFP_model_summary2}.}
\label{table:LFP_model_summary1}
\end{table}
%%%%%%%%%%%%%%%%%%%%%%%%%%%%%%%%%%%%%%%%%%%%%%%%%%%%%%%%%%%%%%%%%%%%%%%%%%%%%%%%%%%%%%%%%%%%

%%%%%%%%%%%%%%%%%%%%%%%%%%%%%%%%%%%%%%%%%%%%%%%%%%%%%%%%%%%%%%%%%%
% TABLE S3 - \label{table:LFP_model_summary2}
%%%%%%%%%%%%%%%%%%%%%%%%%%%%%%%%%%%%%%%%%%%%%%%%%%%%%%%%
\clearpage
\begin{table}[ht!]
    %% Synapse
      \begin{tabularx}{\linewidth}{|p{0.2\linewidth}|X|}
        \hline\modelhdr{2}{D}{Synapse}\\\hline

        \bf Type & current synapse with $\alpha$-function shaped postsynaptic currents (PSCs) \\\hline
        \bf Dynamics & input current: $I(t)=  \sum_i h(t-t_i)$\\

        & with kernel $h(t)$ =
    %    \begin{tabular}{ll}
        $I_0 \frac{t}{\tau} e^{1-(t-t_i)/\tau}\theta(t-t_i)$ \\%& if $t>0$  \\
    %    $0$ & else
    %    \end{tabular}  \\

        & where $t_i$ is the arrival time of presynaptic spike $i$, $\theta$ is Heaviside step function, with time constant $\tau$ and amplitude $I_0$= {$I_e$, $I_i$} for excitatory and inhibitory synapses respectively
       \\\hline
        \bf Parameters & $\tau$, $I_e$, $I_i$ \\\hline

      \end{tabularx}\vspace{\tabspace}

  %% Input
  \begin{tabularx}{\linewidth}{|p{0.2\linewidth}|X|}
    \hline\modelhdr{2}{E}{Input}\\\hline
    \hline\nettypehdr{2}{}{Poissonian input}\\\hline
    \bf Uncorrelated \newline input & each cell receives $n_\text{syn}$ independent excitatory inputs generated by Poissonian point processes with rate $r_{syn}$ \\\hline

    \bf Correlated \newline input & each cell receives $n_\text{syn}$ excitatory inputs drawn (without resampling) from a finite-sized pool (size $n_\text{pool}$) of independent Poissonian point processes with rate $r_\text{syn}$ resulting in correlation $c_\text{in} = n_\text{syn}/n_\text{pool}$ between total input current of different cells \\\hline
    \bf Synapse \newline placement & synapses placed on dendrites in certain layers depending on synaptic input region (apical/homogeneous/basal) and cell type:\\
    & \begin{tabular}{c|c|c|c}
        \textbf{cell type} & \textbf{apical} & \textbf{homogeneous} & \textbf{basal} \\
        \hline
        L3 & upper half of L23 & L1 and L23 & lower half of L23 \\
        \hline
        L4 & - & L4 & - \\
        \hline
        L5 & L1 and L23 & all layers & L5 and L6 \\

    \end{tabular} \\
    &  Random placement of synapses within allowed boundaries with uniform density with respect to membrane area (note: $n_{syn}$ is fixed irrespective of input region) \\\hline
            \bf Parameters & $n_\text{syn}$, $r_\text{syn}$, $c_{\text{in}}$\\\hline

    %     \hline\nettypehdr{2}{}{Laminar-network input}\\\hline
    %     \bf Type & excitatory and inhibitory input generated by laminar network model (see Table S\prettyref{table:laminar_network_summary}) \\\hline
    % 
    %     \bf Synapse \newline placement & see Supplemental Procedures \ref{sec:connecting_models} \\\hline     
  \end{tabularx}\vspace{\tabspace}

  %% Measurements
  \begin{tabularx}{\linewidth}{|p{0.2\linewidth}|X|}
    \hline\modelhdr{2}{F}{Measurements}\\\hline
    \hline\nettypehdr{2}{}{Simulated LFP}\\\hline
    \bf Type & extracellular field potentials (representing the LFP) calculated using the line-source method \\\hline
    \bf Assumptions & extracellular medium as assumed to be purely resistive (non-capacitive) and inifitely-volumed with extracellular conductivity $\sigma_\text{cond}$ \\\hline

    \bf Electrode \newline properties & ideal (non-filtering) point electrode placed either in the center ($r$=0) of the cylindrical geometry, or offset by some distance; at depths corresponding to depth of somata (see Table S1:Population). \\\hline

    \bf Parameters & $\sigma_\text{cond}$ \\\hline

    % \hline\nettypehdr{2}{}{Synaptic input current}\\\hline
    % 
    % \bf Type & \emph{for laminar-network input}: total input current to each cell calculated by summing over input currents to all synapses of specific cell (for use in calculation of $c_{\xi}$)\\\hline    
%
  \end{tabularx}\vspace{\tabspace}
\caption{Summary of the population model used for LFP simulations. Continued from Table \prettyref{table:LFP_model_summary1}}
\label{table:LFP_model_summary2}
\end{table}
%%%%%%%%%%%%%%%%%%%%%%%%%%%%%%%%%%%%%%%%%%%%%%%%%%%%%%%%

%%%%%%%%%%%%%%%%%%%%%%%%%%%%%%%%%%%%%%%%%%%%%%%%%%%%%%%%%%%%%%%%%%
% TABLE S4 - \label{table:LFP_model_params}
%%%%%%%%%%%%%%%%%%%%%%%%%%%%%%%%%%%%%%%%%%%%%%%%%%%%%%%%
%%Parameter for population simulations %%%%%%%%%%%%%%%%%%%%%%%%%%%%%%%%%%%%%

\clearpage
\begin{table}[ht!]
  \def\tabspace{0.5ex}
  %% population parameters
  \begin{tabularx}{\textwidth}{|p{0.15\linewidth}|p{0.5\linewidth}|X|}
    \hline\parameterhdr{3}{A}{Population}\\\hline
    \bf Name & \bf Description & \bf Value\\\hline
    $N$ & number of cells& 10000 \\
    $R$ & population radius & 1000 $\mu$m \\\hline
        \end{tabularx}\vspace{\tabspace}

    \begin{tabularx}{\textwidth}{|p{0.15\linewidth}|X|}
    \hline
   $z_k$ and layer boundaries & layer boundaries and soma positions (in units of $\mu$m in relation to cortical surface at 0 $\mu$m), derived from  Stepanyants et al., Cereb Cortex 2008, 18(1):13--28:
%   \cite{Stepanyants:2008p8003}: \newline

   \begin{tabular}{c|c|c|c}
            \textbf{layer} & \textbf{upper boundary} & \textbf{soma depth $z_k$} & \textbf{lower boundary} \\\hline
            L1 & 0 & - & -81.6 \\
            L2/3 & -81.6 & -334.3 & -587.1 \\
            L4 & -587.1 & -754.6 & -922.2 \\
            L5 & -922.2  & -1021.1 ($^1$)&  -1170.0\\
            L6 & -1170.0 & - & -1491.7\\

        \end{tabular} \\\hline
  \end{tabularx}\vspace{\tabspace}

  %% neuron parameters
  \begin{tabularx}{\textwidth}{|p{0.15\linewidth}|p{0.5\linewidth}|X|}
    \hline\parameterhdr{3}{B}{Neuron}\\\hline
    \bf Name & \bf Description & \bf Value\\\hline
    $R_\text{m}$ & specific membrane resistance & 30 k$\Omega\cdot$cm$^2$ \\
    $R_\text{a}$ & specific axial resistance & 150 $\Omega\cdot$cm \\
    $C_\text{m}$ & specific membrane capacitance & 1.0 $\mu$F/cm$^2$ \\
    \hline
  \end{tabularx}\vspace{\tabspace}

  %% synapse parameters
  \begin{tabularx}{\textwidth}{|p{0.15\linewidth}|p{0.5\linewidth}|X|}
    \hline\parameterhdr{3}{C}{Synapse}\\\hline
    \bf Name & \bf Description & \bf Value\\\hline
    $\tau$ & synaptic time constant & 0.1 ms \\
    $I_e$ & excitatory current amplitude & 50 pA \\
    \hline
  \end{tabularx}\vspace{\tabspace}

  %% input parameters
  \begin{tabularx}{\textwidth}{|p{0.15\linewidth}|p{0.5\linewidth}|X|}
    \hline\parameterhdr{3}{D}{Input}\\\hline
    \bf Name & \bf Description & \bf Value\\\hline
    $n_\text{syn}$ & number of synapses & 1000 \\
    $r_\text{syn}$ & synaptic input rate & 5 spikes/s \\
    $c_\text{in}$ & pairwise input correlation between cells & \{0.01,0.1,1.0\} 
    %Additionally \{0.0002, 0.001 \} for L5 apical
\\
    \hline
  \end{tabularx}\vspace{\tabspace}

  %% measurement parameters
  \begin{tabularx}{\textwidth}{|p{0.15\linewidth}|p{0.5\linewidth}|X|}
    \hline\parameterhdr{3}{E}{Measurements}\\\hline
    \bf Name & \bf Description & \bf Value\\\hline
    $\sigma_\text{cond}$ &  extracellular conductivity  & 0.3 S/m \\
    \hline
  \end{tabularx}\vspace{\tabspace}

  %% simulation parameters
  \begin{tabularx}{\textwidth}{|p{0.15\linewidth}|p{0.5\linewidth}|X|}
    \hline\parameterhdr{3}{F}{Simulation}\\\hline
    \bf Name & \bf Description & \bf Value\\\hline
    dt$_\text{sim}$ &  time resolution in simulation  &  1/64 ms \\
    dt$_\text{data}$ &  time resolution of data &  1.0 ms \\
    T & simulation time & either 10200 ms or 1200~ms (off-center) ($^2$)   \\\hline
  \end{tabularx}\vspace{\tabspace}
($^1$) the somata of the L5 cells were adjusted 25 $\mu$m upwards compared to midpoint of L5 so that apical dendrites reach L1 \newline
($^2$) the first 200 ms were discarded to avoid upstart effects

\caption{Parameters of the population model used for LFP simulations}
\label{table:LFP_model_params}
\end{table}

%%%%%%%%%%%%%%%%%%%%%%%%%%%%%%%%%%%%%%%%%%%%%%%%%%%%%%%%%%%%%%%%%%%%%%%%%%%%%%%

%\section*{References}
% The bibtex filename
% NO WHITESPACE if multiple files
\clearpage
\bibliography{biblio,biblio_Gaute}

\end{document}